\documentclass[review]{elsarticle}

\usepackage{lineno,hyperref}
\usepackage{multirow}
\usepackage{graphicx}
\usepackage{adjustbox}
\usepackage{rotating}
\usepackage{array}
\usepackage{diagbox}
\usepackage{url}

\usepackage{xcolor}
\usepackage{tikz}
\def\checkmark{\tikz\fill[scale=0.4](0,.35) -- (.25,0) -- (1,.7) -- (.25,.15) -- cycle;} 

\journal{Vehicular Communications}

\begin{document}

\begin{frontmatter}

\title{A survey on road safety and traffic efficiency vehicular applications based on C-V2X technologies}

\tnotetext[mytitlenote]{\textbf{Published as: Ignacio Soto, Maria Calderon, Oscar Amador, Manuel Urue\~na, A survey on road safety and traffic efficiency vehicular applications based on C-V2X technologies, Vehicular Communications, Volume 33, 2022, 100428.  The final version of record is available at \href{https://doi.org/10.1016/j.vehcom.2021.100428}{ https://doi.org/10.1016/j.vehcom.2021.100428.}}}
\tnotetext[mytitlenote]{This work was partially supported by the Agencia Estatal de Investigaci\'on (AEI, Spain) through the ACHILLES project (PID2019-104207RB-I00/AEI/10.13039/501100011033) and by the Madrid Government (Comunidad de Madrid-Spain) under the Multiannual Agreement with UC3M in the line of Excellence of University Professors (EPUC3M21), and in the context of the V PRICIT (Regional Programme of Research and Technological Innovation).}


\author[UC3M]{Ignacio Soto\corref{mycorrespondingauthor}}
\cortext[mycorrespondingauthor]{Corresponding author, current address: Departamento de Ingenier\'{\i}a de Sistemas Telem\'aticos, Universidad Polit\'ecnica de Madrid; 28040 Madrid (Madrid), Spain.}
\ead{ignacio.soto@upm.es}

\author[UC3M]{Maria Calderon}
\ead{maria@it.uc3m.es}

\author[Halmstad]{Oscar Amador}
\ead{oscar.molina@hh.se}

\author[UNIR]{Manuel Urue\~na}
\ead{manuel.uruena@unir.net}

\address[UC3M]{Departamento de Ingenier\'{\i}a Telem\'atica; Universidad Carlos III de Madrid; 28911 Legan\'es (Madrid); Spain}
\address[Halmstad]{School of Information Technology; Halmstad University; Halmstad 30118; Sweden}

\address[UNIR]{Escuela Superior de Ingenieros y Tecnolog\'{\i}a, Universidad Internacional de la Rioja, 26006 Logro\~no, Spain}

\begin{abstract}
In recent years, the use of cellular network technologies to provide communication-based applications to vehicles has received considerable attention. 3GPP, the standardization body responsible for cellular networks specifications, is developing technologies to meet the requirements of vehicular communication applications, and the research community is testing and validating the ability of those technologies to implement different applications. This survey presents the body of work dealing with the use of cellular technologies to implement communication-based applications for the connected vehicle. We focus on basic and advanced road safety and traffic efficiency applications, which are critically important for the future of vehicular networks. We start by describing the different cellular-related technologies that have a role to play in providing services to the connected vehicle, propose a classification of types of communication used in vehicular applications, and then apply this classification to organize and present recent research work on the topic. Finally, we identify the main challenges in the use of cellular technologies to develop applications for the connected vehicle. 
\end{abstract}


\end{frontmatter}


\section{Introduction}

There are strong incentives to provide vehicles with communication-based applications to increase the safety of their occupants and other road users, as well as to improve the efficiency in the use of fuel and road infrastructures. Different standardization organizations have worked, and continue working, in developing solutions for the connected vehicle to implement the vision of a future Cooperative, Connected, and Automated Mobility (CCAM). Key results of these efforts are the Intelligent Transport System (ITS) protocol stack \cite{ETSI2010} with ITS-G5 \cite{ETSI2020c} access specified by European Telecommunications Standards Institute (ETSI), and the Dedicated Short Range Communications (DSRC) protocol stack \cite{Kenney2011} developed by SAE International. These protocol stacks enable the communication among vehicles and between vehicles and nodes in the infrastructure (RSUs or Road-Side units).

Both the ETSI ITS with ITS-G5 and the DSRC protocol stacks are based, for their access layers, on IEEE 802.11p (IEEE 802.11 OCB --- Outside the Context of a Basic service set), a mode of operation of the IEEE 802.11 standard \cite{IEEE2020} adapted to the needs of \textcolor{black}{vehicular communications}. For more than 10 years, numerous research efforts have been devoted to the use of this technology to provide services to the connected vehicle, with also a large number of pilot deployments on open roads \cite{InterCor2019} \cite{USDT2021}. Spectrum in the 5.9 GHz band has been allocated to ITS services in different regions of the world. However, the commercial deployment of this technology has not progressed as expected. 

Meanwhile, the 3rd Generation Partnership Project (3GPP) \cite{3gpp} started to develop solutions to enable the use of cellular technologies in the access layer of connected vehicle communications, an effort jointly known under the name of \mbox{C-V2X} (Cellular Vehicle-to-Everything), where "everything" (i.e., the X) can be another vehicle, a pedestrian, the road infrastructure or a server on the network. 3GPP Release 14 (from mid-2017) introduced, for the first time, D2D (direct) communications between terminals specifically intended for vehicle-to-vehicle communications. These efforts have resulted in a shift to focus on C-V2X technologies to support vehicular communications, for example reflected in the \textcolor{black}{United States Federal Communications Commission} (FCC) decision to change the use of the reserved spectrum for intelligent transportation system operations in the 5.9 GHz band from DSRC to C-V2X~\cite{FCC2020}.

The 3GPP is in a constant process of improving the technology, considering requirements for increasingly advanced services for the connected vehicle. The research community has also focused on analyzing and improving the \mbox{C-V2X} technologies to support the connected vehicle, and comparing them to IEEE 802.11p based approaches. 

There are three main categories for connected vehicle applications: 
\begin{enumerate}
\item Road safety: applications that assist in the protection of vehicles, their occupants, and other road users. A special case is the protection of Vulnerable Road Users (VRUs), i.e., users not protected by a vehicle body, such as pedestrians, cyclists and motorcyclists.

\item Traffic efficiency: applications that improve efficiency in the use of vehicles. These applications can help save fuel, travel time, or make a better use of roads to serve more users with the same infrastructure. In some cases, there is an overlap between this category and the previous one: for example, an application that facilitates vehicles merging onto a road improves both safety and efficiency.
\item Others: applications such as, for example, convenience applications, and access to information or entertainment (infotainment applications). 
\end{enumerate}

In this article we review the research on using C-V2X communications to provide road safety and traffic efficiency applications to the connected vehicle, although other applications are also contemplated when some work in the literature analyzes them co-existing with road safety and traffic efficiency applications.

\textcolor{black}{\subsection{Related work}}
\textcolor{black}{
In recent years, a number of surveys have been published in different areas related to the support of connected vehicle applications, using different access technologies including C-V2X communication technologies.
Araniti et al. \cite{Araniti2013} in an early paper investigated the usability of Long Term Evolution (LTE) to support vehicular applications to face the drawbacks of IEEE 802.11p technology. Masini et al. \cite{Masini2018} reviewed various access technologies for Vehicle-to-Vehicle (V2V) communications to enable vehicular sensor networks, including economic motivations and business models.
The work of Singh et al. \cite{Singh2019} had a much wider scope, reviewing in a tutorial style different communication technologies; protocol stacks specified in USA, Japan and Europe; applications; standardization efforts; and description of related research projects. 
Ahangar et al. \cite{Ahangar2021} investigated the autonomous vehicle use case and the service provided by different vehicular communications technologies with different coverage ranges (from Zigbee to 5G-New Radio) paying attention to latency requirements. 
Chen et al. \cite{Chen2020} studied C-V2X as enabling technologies to meet the requirements of autonomous driving, presenting the technical evolution path from LTE Vehicle-to-Everything (LTE-V2X) to New Radio-V2X (NR-V2X) and highlighting field trials and developments in China. 
The paper of Gyawali et al. \cite{Gyawali2021} studied the challenges for LTE and New Radio (NR) to support Vehicle-to-Everything (V2X) communications in general, without discussing any particular application, especially concentrating physical layer structure, synchronization and security. 
The work of Bazzi et al. \cite{Bazzi2021} provided a survey about different aspects of the sidelink interface design taking into account the evolutionary path of sidelink interface in 3GPP. Interestingly, the paper includes a discussion on congestion control mechanisms for sidelink interface.
Alalewi et al. \cite{Alalewi2021} focused on the 5G enabling  technologies and their potential applicability to relevant V2X use cases such as platooning or remote driving, specially concentrating on the QoS requirements of the analysed use cases. It includes an historical review of vehicular communications, considering projects, technologies and standards.} The topics covered by Singh et al. 
The work of Jeong et al. \cite{Jeong2021b} surveys IP-based vehicular networking, and the analyzed issues are related with the particularities of using IP transport on vehicular networks: IP-based applications, IP address autoconfiguration, IP mobility management, and IP security. 
Finally, Lee et al. \cite{Lee2020} provided an interesting survey about the connected vehicle applications and their benefits, but in a communication technologies-agnostic way. 

\textcolor{black}{
Other related surveys focused on the use of IEEE 802.11p technology to provide vehicular communications, as the survey of Toor et al. \cite{Toor2008} where the main issues addressed were Internet connectivity, data dissemination, and routing over IEEE 802.11p, or the work of Jeong at al, \cite{Jeong2021} which surveyed systems, protocols, applications, and security for safe and efficient driving, but most surveyed cases use IEEE 802.11p as access technology, and it does not analyze how the access technology performs to implement these use cases. 
There are other surveys that covered different aspects related to vehicular networking, such as the comparison of Wi-Fi and sidelink cellular technologies for communicating vehicles \cite{Bazzi2019}, Internet of Vehicles based on C-V2X technologies \cite{Zhou2020} \cite{Storck2020}, radio resource allocation in C-V2X technologies \cite{Allouch2020}, \cite{Le2021} \cite{Masmoudi2019}, security \cite{Muhammad2018}, or machine learning applied to vehicular communications \cite{Tang2020}.
}

\textcolor{black}{
Table~\ref{table:othersurveys} shows a comparison of this survey and other recent related surveys. The comparison is based on the following aspects: C-V2X technologies covered; whether they include a description (tutorial style) of C-V2X and related technologies; whether they include an analysis by type of end-point (vehicle, pedestrian, infrastructure, or network) and characteristics of messages exchanged; whether it classifies the surveyed papers by experimentation methodologies; whether it includes an analysis of communication services to support different types of applications; use cases; standardization; and other related technologies.}

\begin{table}[tb!] \footnotesize
\begin{center}
\begin{adjustbox}{max width=1.1\textwidth}
\begin{tabular}{|p{2cm}|p{1.4cm}|p{1.9cm}|p{2cm}|p{1.8cm}|p{2cm}|p{1.5cm}|c|p{1.4cm}|}
\hline
\multirow{7}{*}{\textbf{Reference}} & \centering \multirow{7}{*}{\textbf{C-V2X}} & \vspace{0.5cm} \centering \textbf{Tutorial of \mbox{C-V2X} and related technologies} & \centering \textbf{Analysis by type of end-point and characteristics of the messages exchanged} & \centering \textbf{Analysis of \mbox{C-V2X} communication services to support applications} & \vspace{0.5cm} \centering \textbf{Classification by experimentation methodology} & \vspace{1cm} \centering \textbf{V2X Use Cases} & \multirow{7}{*}{\textbf{Standardization}} & \centering \arraybackslash \multirow{7}{*}{\textbf{Other}} \\ \hline 
\vspace{-0.1cm} Araniti et al., 2013 \cite{Araniti2013} & \flushleft \vspace{-0.5cm} LTE Uu; LTE PC5; MBMS & \centering \multirow{3}{*}{X} & \centering \multirow{3}{*}{X} & \centering \multirow{3}{*}{\checkmark} & \centering \multirow{3}{*}{X} & \centering \multirow{3}{*}{X} & \centering \multirow{3}{*}{\checkmark} & \centering \arraybackslash \multirow{3}{*}{X} \\ \hline
\vspace{-0.1cm} Masini et al., 2018 \cite{Masini2018} & \vspace{-0.3cm} \flushleft LTE PC5; NR PC5 & \centering \multirow{3}{*}{\checkmark} & \centering \multirow{3}{*}{X} & \centering \multirow{3}{*}{X} & \centering \multirow{3}{*}{X} & Vehicular sensor networks & \centering \multirow{3}{*}{\checkmark} & \centering \arraybackslash \multirow{3}{*}{X} \\ \hline
\vspace{-0.1cm} Singh et al., 2019 \cite{Singh2019} & \vspace{-0.3cm} \flushleft LTE Uu; LTE PC5 & \centering \multirow{3}{*}{\checkmark} & \centering \multirow{3}{*}{X} & \centering \multirow{3}{*}{X} & \centering \multirow{3}{*}{X} & \centering \multirow{3}{*}{X} & \centering \multirow{3}{*}{\checkmark} & \centering \arraybackslash \multirow{3}{*}{X} \\ \hline
\vspace{0.15cm} Chen et al., 2020 
\cite{Chen2020} & \vspace{-0.5cm} \flushleft LTE Uu; LTE PC5; NR Uu; NR PC5 & \centering \multirow{4}{*}{\checkmark} & \centering \multirow{4}{*}{X} & \centering \multirow{4}{*}{X} & \centering \multirow{4}{*}{X} & \vspace{0.15cm} Autonomous vehicle & \centering \multirow{4}{*}{\checkmark} & \centering \arraybackslash \multirow{4}{*}{MEC} \\ \hline
\vspace{1.1cm} Bazzi et al., 2021 \cite{Bazzi2021} & \vspace{1cm} \flushleft LTE PC5; NR PC5 & \centering \multirow{8}{*}{\checkmark} & \centering \multirow{8}{*}{X} & \centering \multirow{8}{*}{X} & \centering \multirow{8}{*}{\checkmark} & Vehicle platooning; remote driving; collective perception; advance driving & \centering \multirow{8}{*}{\checkmark} & \centering \arraybackslash \multirow{8}{*}{X} \\ \hline
\vspace{0.15cm} Gyawali et al., 2021 \cite{Gyawali2021} & \vspace{-0.5cm} \flushleft LTE Uu; LTE PC5; NR; MBMS & \centering \multirow{4}{*}{\checkmark} & \centering \multirow{4}{*}{X} & \centering \multirow{4}{*}{X} & \centering \multirow{4}{*}{X} & \centering \multirow{4}{*}{X} & \centering \multirow{4}{*}{\checkmark} & \vspace{-0.05cm} SDN; Network slicing; MEC \\ \hline
\vspace{-0.1cm} Ahangar et al., 2021 \cite{Ahangar2021} & \flushleft \vspace{-0.5cm}  LTE Uu; LTE PC5; NR & \centering \multirow{3}{*}{X} & \centering \multirow{3}{*}{X} & \centering \multirow{3}{*}{X} & \centering \multirow{3}{*}{X} &  \vspace{0.001cm} Autonomous vehicle & \centering \multirow{3}{*}{\checkmark} & \centering \arraybackslash \multirow{3}{*}{X} \\ \hline
\vspace{1.1cm} Alalewi et al., 2021 \cite{Alalewi2021} & \vspace{0.95cm} \flushleft NR Uu; NR PC5 & \centering \multirow{8}{*}{X} & \centering \multirow{8}{*}{X} & \centering \multirow{8}{*}{\checkmark} & \centering \multirow{8}{*}{X} & Vehicle platooning; remote driving; collective perception; advance driving & \centering \multirow{8}{*}{\checkmark} & \vspace{0.35cm} SDN; NFV; Network slicing; MEC \\ \hline
\vspace{0.7cm} Our survey & \vspace{-0.5cm} \flushleft LTE Uu; LTE PC5; NR Uu; NR PC5; eMBMS & \centering \multirow{5}{*}{\checkmark} & \centering \multirow{5}{*}{\checkmark} & \centering \multirow{5}{*}{\checkmark} & \centering \multirow{5}{*}{\checkmark} & \centering \multirow{5}{*}{28} & \centering \multirow{5}{*}{\checkmark} & SDN; NFV; Network slicing; MEC \\ \hline
\end{tabular} 
\end{adjustbox}
\caption{\textcolor{black}{Comparison of this survey with other recent related surveys.}}
\label{table:othersurveys}
\end{center}
\end{table}

\textcolor{black}{\subsection{Current Survey}}
\textcolor{black}{The contribution of our survey focuses on the literature that analyzes the utilization of the different C-V2X technologies to support connected vehicle applications, studying their appropriateness to meet their requirements, their limitations, and exploring how to use and configure the technologies for this purpose. 
To this end, this paper introduces several novel aspects that, as a whole, have not been covered by other surveys in the past:
\begin{enumerate}
     \item We feature a brief tutorial on C-V2X communication technologies, including the latest ones such as 5G NR (New Radio) cellular and NR sidelink.  
     \item We categorize the types of communication and related messages needed to enable different applications for the connected vehicle (cooperative collision warning, see-through or collective perception are some examples of the 28 specific applications considered in the survey). Upon this categorization, we extend the 3GPP classification of types of V2X communications that only considers the type of end-point (e.g., vehicle, pedestrian, infrastructure) to take also into account the characteristics of the exchanged messages (e.g., periodic, event-driven, relayed by RSU, or multi-hop). 
     \item We explore the recent literature on proposals of C-V2X related solutions to provide the communications services (types of communication and associated messages) required by connected vehicle applications.
     \item We categorize surveyed papers by the experimentation methodology they use (e.g., focused simulations, system-level simulations, real equipment tests, field tests and demonstrations). 
     \item We present several challenges for C-V2X applications research and deployment.
\end{enumerate}
}

\begin{figure}[t]
	\centering
	\includegraphics[width=\textwidth]{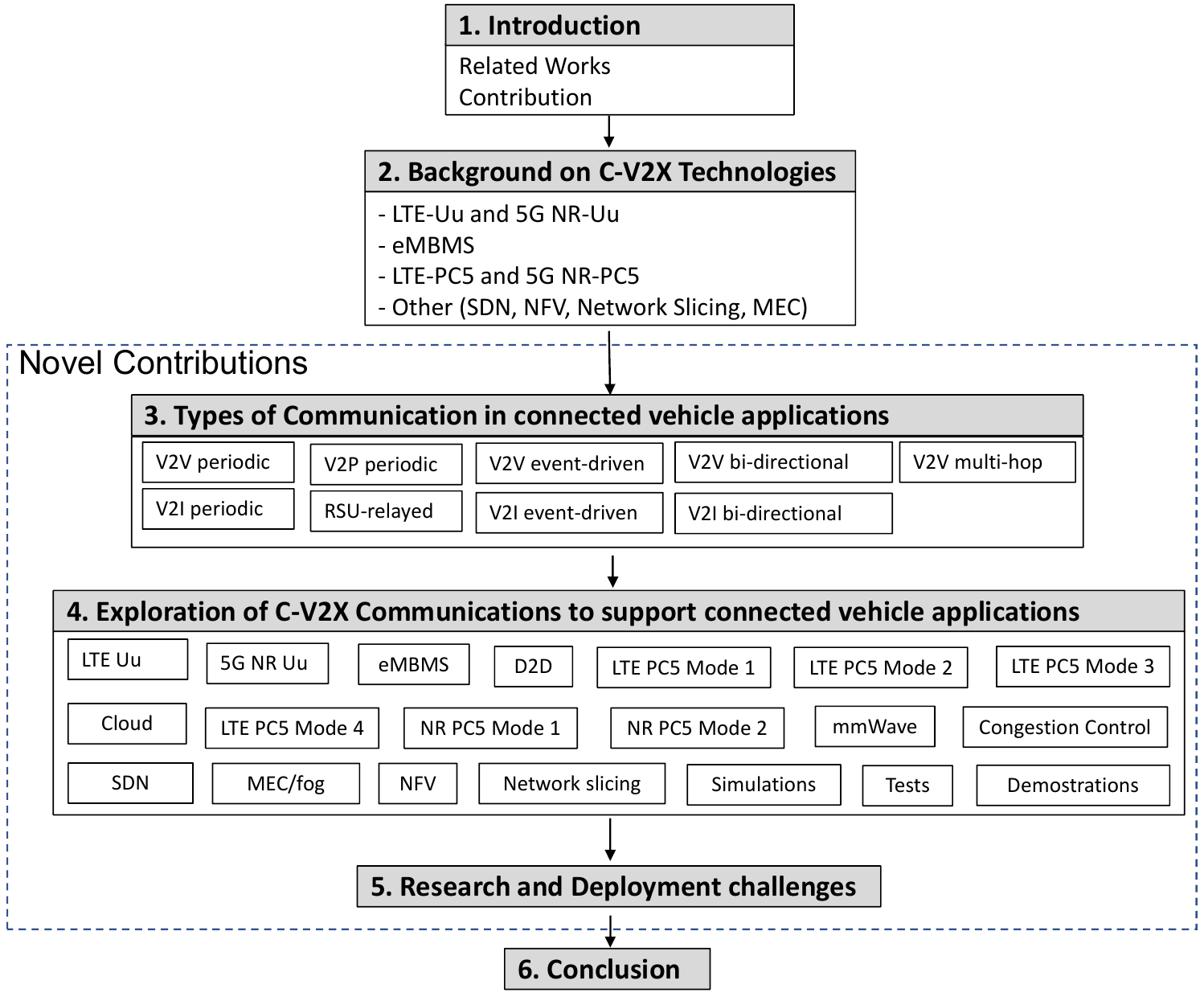}
	\textcolor{black}{\caption{Structure of this survey}}
	\label{fig:flowchart-v2}
\end{figure}

The rest of the article is organized as follows: section~\ref{sec:technologies} briefly reviews the C-V2X communication technologies; section~\ref{sec:types} categorizes the different types of communication to support connected vehicle applications; section~\ref{sec:exploration} organizes and summarizes, with the previous classification scheme, the contributions of the literature studying C-V2X based applications; section~\ref{sec:challenges} identifies the main future challenges for the research and development of C-V2X applications; and finally, section~\ref{sec:conclusion} summarizes the conclusions of our survey. \textcolor{black}{The organization of the survey is shown in Figure~\ref{fig:flowchart-v2}. The objective is to make it easier for readers to understand the topics covered and its general structure.}

\section{Background: C-V2X communication technologies}
\label{sec:technologies}

The 3GPP \cite{3gpp} develops specifications for cellular network technologies. 
The specifications are organized in \textit{Releases}. The most recent, as of this writing, is \textit{Release 17}, scheduled for 2022 (although some work has already been started on Release 18). Since \textit{Release 14} (issued in 2017), the 3GPP has 
been developing communication technologies tailored for the connected vehicle, 
under the name \textcolor{black}{Vehicle-to-Everything} (V2X) technologies or Cellular-V2X (C-V2X), but even before that milestone, researchers had been exploring the use of cellular communication technologies to support applications for the connected vehicle as an alternative to the previous solutions based on IEEE 802.11p communications.  

\begin{figure}[t]
	\centering
	\includegraphics[width=\textwidth]{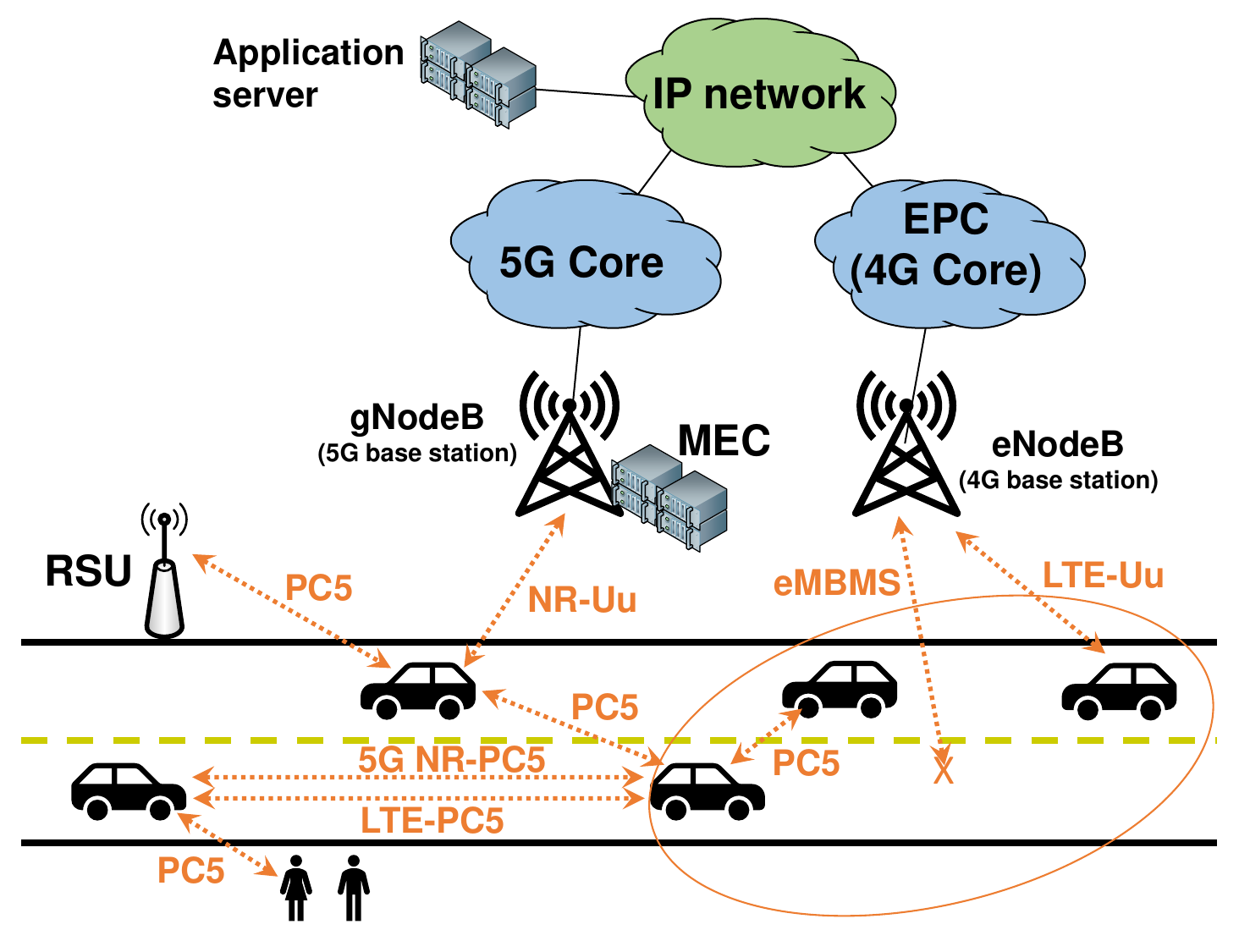}
	\caption{C-V2X communication interfaces}
	\label{fig:interfaces}
\end{figure}

In this section we briefly present the different cellular communication technologies, and related ones that can be used to support applications for the connected vehicle. Some of the communication interfaces and technologies are represented in Figure~\ref{fig:interfaces}.

\subsection{LTE Uu and 5G NR Uu}

Uu is the traditional radio interface of the cellular communication network, which allows terminals to access the network through a base station in the infrastructure. LTE Uu is the 4G version of this interface, while 5G NR (New Radio) Uu is the 5G version of this interface (included initially in Release 14, and in continuous development in later Releases). 5G NR Uu allows the use of new frequency bands (including mmWave), and it enables higher throughput, less latency, higher reliability, and improved spectrum efficiency.

\subsection{eMBMS}

eMBMS (evolved Multimedia Broadcast Multicast Service) provides point-to-multipoint communications through the Uu interface of 3GPP cellular networks \cite{Guo2018} \cite{Gomez-Barquero2020}. The initial  method for resource allocation in eMBMS is called MBSFN (Multicast Broadcast Single Frequency Network), and was designed for video channel broadcasting. MBSFN is based on a strict scheduling of services, synchronized transmissions to a group of cells, and dedicated physical channels for broadcast traffic. An alternative method for resource allocation of broadcast traffic was introduced in Release 13, namely SC-PTM (Single-Cell Point-To-Multipoint) which enables delivering services to one cell and does not require dedicated channels for broadcast traffic, achieving more flexibility and less latency. SC-PTM has been, therefore, the method chosen for V2X communications \cite{Sacristan2018}. Further improvements for broadcast communications for V2X applications are expected in Release 17.

\subsection{LTE PC5 and 5G NR PC5}

Sidelink communications (direct communication between terminals) were introduced in 3GPP specifications in Release 12 (2015) but focusing on supporting proximity services \textcolor{black}{\cite{3GPP2015b} \cite{Lin2014}}, for example, for communications of public safety operatives. PC5 is the name in 3GPP networks for the sidelink interface. Release 12 LTE PC5 interface has two modes of operation \textcolor{black}{\cite{3GPP2014}}: mode 1 in which the resources are assigned from the network; and therefore, network coverage is required for operation, and mode 2 in which resources are autonomously allocated by each terminal; and therefore, network coverage is not needed for operation. 

3GPP Release 14 (2017) was the first to introduce sidelink communications explicitly designed for V2X communications \textcolor{black}{\cite{3GPP2016} \cite{3GPP2017} \cite{3GPP2017b} \cite{3GPP2017c}} \cite{Molina-Masegosa2017}. In particular, two modes of operation were introduced:

\begin{enumerate}
\item Release 14 LTE PC5 mode 3: supports direct broadcast communications between vehicles, but the resource allocation for the communication is from the network, which requires vehicles to be in the radio coverage of a cellular base station.
\item Release 14 LTE PC5 mode 4: supports direct broadcast communications between vehicles, but the resource allocation for the communication is done autonomously by each vehicle. Therefore, vehicles do not need to be in the radio coverage of a cellular base station. The autonomous allocation of resources is done with an algorithm called S-SPS ("Sensing-based SemiPersistent Scheduling"). 
\end{enumerate}  

A high level description of how the S-SPS algorithm works is as follows (for a more detailed description, see \cite{Molina-Masegosa2017} or \cite{Bazzi2020}):

\begin{enumerate}
\item The objective for a vehicle is to assign to itself radio resources to be able to broadcast a message on the sidelink interface. Radio resources are, basically, a slot in time and a chunk of bandwidth that can allow sending the message.
\item The vehicle chooses a selection window, where there are resources that could allow to transmit the message while fulfilling its timing requirements.
\item The vehicle uses a sensing window, previous to the selection window, during which the vehicle checks the utilization of the medium (a vehicle senses the media while not transmitting). There are two ways to detect a resource as busy: the received power or when a future resource is announced as reserved in the control information sent with a message (see below).  
\item Based on the feedback obtained in the sensing window, the vehicle can identify free candidate resources in the selection window, which could be used by the vehicle to send its message. From among them, one is randomly selected from a subset of the best candidates.
\item The resource selection is kept for a number of repetitions. The time between repetitions and when resources are re-selected is configurable. A re-selection can also be triggered if needed to fulfill latency requirements.
\end{enumerate}

Note that the algorithm relies on a semi-persistent scheduling of resources, which allows to discover the future utilization of resources from the information captured on the sensing phase. Vehicles announce, as part of the control block sent with a message, if they intend to use a future slot and the time period until that slot (from a fixed set of resource reservation intervals). This works well with messages sent at fix periodic intervals and with fixed sizes \cite{Bartoletti2021}. Some limitations of this scheduling mechanism are the half-duplex problem (a vehicle transmitting using some frequency resources at one time slot cannot hear vehicles transmitting at the same time slot even in a different frequency), and bursts of collisions (if two vehicles choose the same resources and period, they will not be able to detect their collisions). The semi-persistent allocation makes these limitations worse, since two vehicles that are transmitting repeatedly at the same time become invisible to each other. \textcolor{black}{The work in \cite{Gonzalez-Martin2019} developed analytical models to study the performance of Release 14 LTE PC5 mode 4}.

A new version of D2D PC5 modes for C-V2X communications was developed in Release 16 (2020) \cite{Garcia2021} \cite{Ali2021} \cite{3GPP2020b} \cite{3GPP2021} \cite{3GPP2021b}. These new modes are based on 5G NR, and they are not meant to replace the LTE PC5 interface modes but to complement them --- LTE PC5 offers basic services in the ITS reserved band at 5.9~GHz, while NR PC5 provides advanced services with higher requirements and, potentially, using alternative spectrum at licensed and unlicensed bands. The two NR PC5 modes are:
\begin{enumerate}
\item Release 16 NR PC5 mode 1: supports direct unicast, groupcast or broadcast communications between vehicles, but the resource allocation for the communication is done by the network, which requires vehicles to be under the radio coverage of a cellular base station.
\item Release 16 NR PC5 mode 2: supports direct unicast, groupcast, or broadcast communications between vehicles, but the resource allocation is done autonomously by each vehicle. Therefore, the vehicles do not need to be under the radio coverage of a cellular base station. 
\end{enumerate} 

The NR PC5 mode 2 defines two ways for the autonomous allocation of resources. In one, resources are allocated for an individual message and their retransmissions (when HARQ, --- Hybrid Automatic Repeat Request --- is used to improve reliability), which is appropriate to send event-driven messages, a use case that was not well addressed with the S-SPS algorithm. The second one is another semi-persistent allocation algorithm, a modified version of the S-SPS algorithm, which is appropriate to reserve resources for sending periodic messages.

In more detail, the improvements in the NR PC5 mode 2 mechanism for resource allocation are as follows: (1) the ability to deal more efficiently with event-driven messages by means of an improved HARQ mechanism that adds, for unicast and groupcast communications, no-blind HARQ to the blind HARQ\footnote{Without going into details, blind HARQ consists on retransmiting a packet preventively to improve reliability; while no-blind HARQ uses feedback from receivers, in unicast and groupcast communications, to decide if retransmissions are needed.}  used in LTE PC5 mode 4; (2) more flexibility on the possible resource reservation intervals that a vehicle can choose for its periodic messages (although still from a limited set of options); (3) smaller minimum latency (1 ms); (4) a mechanism to exclude candidate resources that can collide with reservations made when the vehicle cannot hear them (half-duplex issue); (5) a re-evaluation mechanism that allows a vehicle to check that its reserved resources have not been reserved by other vehicles after its initial selection; and (6) a preemption mechanism so that a vehicle has to withdraw its reserved resources if, after a re-evaluation, finds that a vehicle with higher priority traffic is going to use them. These improvements provide more flexibility and adaptability to allocate resources meeting the different requirements of the traffic generated in vehicles, and reducing collisions due to conflicts among different allocations.

\subsection{Additional technologies}

The efficient and flexible support of ultra-reliable low-latency communications for V2X services provided through the cellular network infrastructure requires the combination of various technologies, of which we highlight the following:

\begin{itemize}
\item \textbf{Software Defined Networking (SDN)} \cite{Xia2015}\cite{Stallings2015} consists in separating the control plane and the data plane in the network. The control function is centralized in nodes called network controllers, which offer programming interfaces that allow SDN applications to program the network. Using SDN, the network becomes much more flexible and adaptable to dynamic requirements.
\item \textbf{Network Function Virtualization (NFV)} \cite{Mijumbi2016} \cite{Stallings2015} moves network functions from dedicated hardware appliances to virtualized functions running in general purpose hardware. This provides greater flexibility in when, where, and how to deploy functions (networking services) that are needed in the network.
\item \textbf{Network Slicing} \cite{Campolo2017b} \cite{Afolabi2018} allows to offer different network services using the same physical infrastructure, so it is possible to provide a customized network service to meet application needs. Through network slicing, the network can guarantee low latency requirements to some applications while serving high bandwidth demands from other applications, all without wasting network resources.   
\item \textbf{Muti-access Edge Computing (MEC)} \cite{Spinelli2021} \cite{ETSI2020b} deploys IT infrastructure with virtualization capabilities at the edge of the network. This infrastructure allows to deploy network functions and applications at the edge of the network. For example, a virtual core network and applications can be deployed in a base station, so the application can offer a low latency service to terminals in the corresponding area (i.e., the cell), because the traffic does not have to go through the physical core of the network and then through Internet to reach the application server. MEC specifications include a set of APIs (application programming interfaces) to manage the applications as well as to allow applications to access information in the network that can be useful (such as the quality of the radio communication with a terminal or the terminals present in a cell).   
\end{itemize}

All these technologies are related and must be strongly integrated in the network to meet the requirements of vehicular communication applications.

\section{Types of communication used in applications for the connected vehicle}
\label{sec:types}

An application for connected vehicles has several components: transfer of information, acquisition
of information from sensors and through communications, processing of information, and human-machine interface aspects. In this paper we focus on the information transfer component. 3GPP identifies the following types of communication involved in V2X applications \cite{3GPP2020} (represented in Figure~\ref{fig:communicationtypes}):

\begin{figure}[t]
	\centering
	\includegraphics[width=\textwidth]{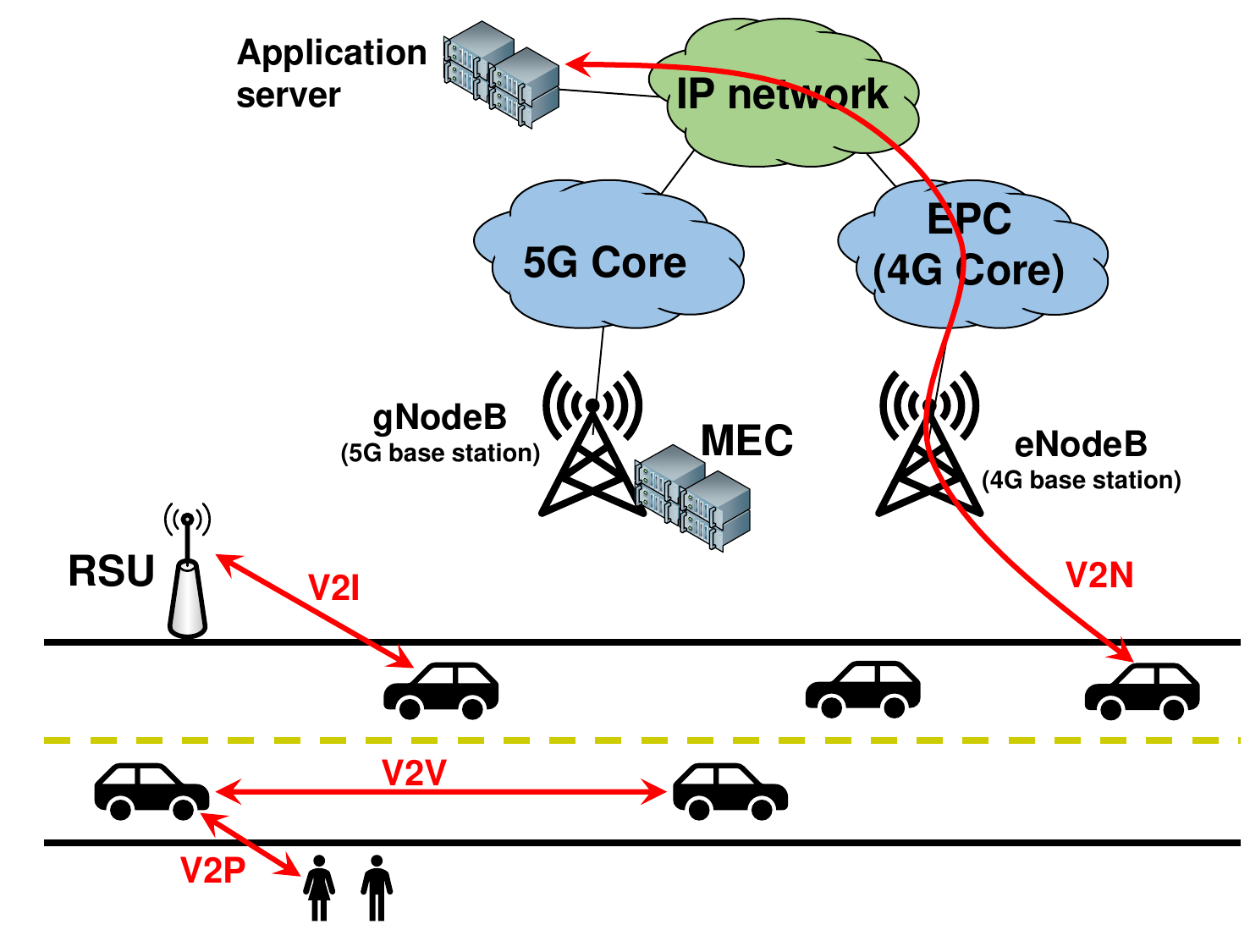}
	\caption{Types of communication in C-V2X}
	\label{fig:communicationtypes}
\end{figure} 

\begin{enumerate}
\item \textbf{V2V --- Vehicle-to-Vehicle} communications in which vehicles send messages to other vehicles. 
\item \textbf{V2P --- Vehicle-to-Pedestrian:} communications in which vehicles send messages to pedestrians, or pedestrians send messages to vehicles. In this case, the communication solutions could be similar to the V2V case, but there can be particularities associated with considerations about pedestrians' devices (e.g., battery availability).
\item \textbf{V2I  --- Vehicle-to-Infrastructure:} communications in which vehicles send messages to the infrastructure, in particular to RSUs (Road Side Units), or RSUs send messages to vehicles. An RSU is a node (e.g., a traffic light or a road signal) located close to a road that can exchange V2X messages with other entities such as vehicles or pedestrians. A local server can be associated with the RSU for message processing.
\item \textbf{V2N --- Vehicle-to-Network:} communications in which vehicle exchanges messages with a V2X application server in the network.
\end{enumerate} 

The difference between V2I and V2N communications is that V2I involves communications with vehicles in the geographical area of the related RSU, while this is not the case with V2N communications in which servers can provide service to different geographical areas. V2I are one-hop communications between RSUs and vehicles, while V2N communications require an IP transport to communicate with a server potentially at multiple hops through the network infrastructure (e.g., involving the use of the core part of the cellular network to communicate with external networks). Even so, the two cases can be difficult to differentiate in some situations, for example when using solutions that move application processing closer to the edge of the network (i.e., MEC). 

In this paper, and based on different reviews of applications for the connected vehicle \cite{CAMP2005} \cite{3GPP2015} \cite{3GPP2018} \cite{Singh2019} \cite{Lee2020} \cite{ETSI2009}, we extend the 3GPP classification of types of V2X communication, not only considering the type of \textcolor{black}{end-point} (e.g., vehicle, pedestrian, infrastructure) but also the characteristics of the messages being exchanged. The considered types of communication are the following:
\begin{enumerate}
\item \textbf{V2V periodic messages:} messages that are periodically sent by a vehicle to all surrounding vehicles, to provide and update information about itself. The period used to send messages is not necessarily fixed, since in many situations the period could be dynamically adjusted to traffic or network conditions. The most significant example is awareness messages, used to inform about the location and kinematic data of the vehicle. Examples of awareness messages are the Cooperative Awareness Message (CAM) defined by ETSI \cite{ETSI2019} and the Basic Safety Message (BSM) defined by SAE International \cite{SAE2020}. The information from awareness messages can be used by other vehicles 
to implement different applications, for example, a warning about a risk of collision, or a cooperative adaptive cruise control that dynamically reduces the speed when the vehicle in front is too close.
\item \textbf{V2V event-driven messages:} messages that are sent by a vehicle to surrounding vehicles to notify an event. A message associated with an event is usually repeated several times over a period of time, but the duration of this time is limited and is related to the duration of the event that triggered the message. An example is a vehicle that detects ice on a road segment and informs surrounding vehicles of the situation. ETSI Decentralized Environmental Notification Messages (DENMs) are an example of event-driven messages. Some V2V applications, such as platooning, use a combination of periodic and event-driven messages.
\item \textbf{V2V \textcolor{black}{bi-directional} communications:} point-to-point \textcolor{black}{bi-directional} communications between two vehicles. An example is a vehicle that wishes to overtake the vehicle in front but has limited visibility, so it requests streaming video of the road ahead from the vehicle in front (see-through application).
\item \textbf{V2V multi-hop messages:} sometimes it is convenient to extend the coverage of V2V or V2I Intelligent Transport System (ITS) messages beyond the radio coverage of the entity sending the message. This can be achieved by other vehicles forwarding the messages, or even, if there are not nearby vehicles, by a vehicle carrying the message and forwarding it later on. For example, a message alerting about an icy road segment could be sent to a geographical area farther from the vehicle detecting it, to give drivers in other vehicles more time to react.   
\item \textbf{V2P periodic messages:} messages that are periodically sent by a vehicle to all surrounding pedestrians, to provide and update information about itself; or messages that are periodically sent to all surrounding vehicles by a pedestrian's device to provide and update information about the pedestrian. In most cases the messages are the same as in V2V periodic messages, only the participants in the communication change. 
\item \textbf{V2I periodic messages:} messages that are sent periodically by an RSU to all vehicles in its coverage area. For example, an RSU can send information about the geometry of an intersection, about traffic signals, or about traffic lights cycles. 
\item \textbf{V2I event-driven messages:} messages that are sent by an RSU to all vehicles in its coverage area, or to particular vehicles in it, to inform about an event; or messages sent by a vehicle to the infrastructure. A message associated with an event is usually sent several times over a period of time, but the duration of this time is limited and related to the duration of the event that triggered the message. An example is an RSU that uses sensors to detect people in a pedestrian crossing and warns approaching vehicles. Another example is an emergency vehicle approaching an intersection that sends a message to an RSU in the intersection (e.g., a traffic light) to request the right-of-way.    
\item \textbf{RSU-relayed messages:} when V2V messages received by RSUs are forwarded to other interest areas to extend their coverage, or are forwarded to a server in the network for further processing. For example, an RSU in an intersection can receive and forward V2V awareness messages to improve coverage in the presence of obstacles; or the RSU could send the messages to a traffic control server, which could use them for traffic monitoring or to execute actions such as controlling traffic lights cycles. This type of communication assumes that messages are not sent directly to the server (which would be a V2N communication), but broadcast in 1 hop by vehicles and received by an RSU.  
\item \textbf{V2N \textcolor{black}{bi-directional} communications:}  point-to-point full duplex communications between a vehicle and a server in the network. This type of communication requires the use of IP transport to reach the server in the network. An example is a vehicle sending information about the current traffic situation or other conditions in the road, and downloading updated dynamic maps. 
\end{enumerate}  

\textcolor{black}{Not all possibilities are present in this classification. For example, some of the message exchange types in the V2V case are not present in other cases. The reason is that we could not find in the literature examples of applications using them. This could be because the particular communication type has not been explored yet in vehicular applications or because the type is not really applicable.}

In Table~\ref{table:commtypes} we present the different types of communication considered in this paper, giving examples of applications, classified by application category (e.g., road safety, traffic efficiency, or convenience). Moreover, the table also includes the papers that are later reviewed and that explore each type of communication  (sometimes a paper appears more than once, if it explores more than one type of communication).

\textcolor{black}{From Table~\ref{table:commtypes}, it can be concluded that most research on C-V2X has been focused on V2V periodic messaging, which is considered the most basic ITS service and present in "Day 1" use cases \cite{C-ITS2016} and pilots \cite{C-ROADS}, together with V2I communications, which are the second area with more research. Moreover, it is worth noting that most of these works are focused on single-hop communications, with few papers considering multi-hop communications or RSU-relayed messages.}

\begin{table}[tb!] \footnotesize
\begin{center}
\begin{adjustbox}{max width=1.1\textwidth}
\begin{tabular}{|l|l|p{6.5cm}|p{3.7cm}|}
\hline
\textbf{V2X Communication Types} & \textbf{Category} & \textbf{Example Applications} & \textbf{Related papers} \\
\hline
\multirow{8}{*}{V2V periodic messages} & \multirow{5}{*}{Road Safety} & 
Cooperative collision warning & \multirow{8}{3.7cm}{\cite{Sun2016} \cite{Bazzi2016} 
\cite{Soleimani2017} \cite{Soleimani2017b} \cite{Gallo2017}
\cite{Peng2017} \cite{Bazzi2017} 
\cite{Kawasaki2017} \cite{Peng2017b} 
\cite{Gharba2017} \cite{Campolo2017} 
\cite{Vukadinovic2018} \cite{Soleimani2018} 
\cite{Wang2018} \cite{Park2018} \cite{Amjad2018} \cite{Sacristan2018} 
\cite{Calabuig2018} \cite{Sacristan2018b}  \cite{Nardini2018} \cite{Molina-Masegosa2018} \cite{Gupta2018} 
\cite{Gupta2019} \cite{Peng2019} \cite{Geng2019} 
\cite{Hegde2019} \cite{Mignardi2019} \cite{Schiegg2019} \cite{Sabeeh2019} \cite{Shimizu2019} \cite{CAMP2019} \cite{5GAA2019} \cite{Llatser2019} \cite{Molina-Masegosa2020} \cite{Hirai2020} \cite{Romeo2020} \cite{Romeo2020b} 
\cite{Niebisch2020} \cite{Shimizu2020} \cite{Choi2020} \cite{Convex2020} \cite{Lekidis2021} \cite{Segata2021} \cite{Hegde2021}} \\ 
& & Intersection movement assistance & \\
& & Slow vehicle warning & \\
& & Cooperative glare reduction & \\ 
& & Collective perception
&  \\ \cline{2-3} 
& \multirow{2}{*}{Traffic Efficiency} & 
Cooperative Adaptive Cruise Control & \\
& & Platooning 
& \\  \cline{2-3} 
& Convenience & Stolen vehicle alert  & \\
\hline
\multirow{5}{*}{V2V event-driven messages}  & \multirow{4}{*}{Road Safety} &  Stationary vehicle warning &
 \multirow{5}{3.5cm}{\cite{Sun2016} \cite{Peng2017} \cite{Geng2019} \cite{Peng2017b}
\cite{Wang2018} \cite{Romeo2020} \cite{Romeo2020b} \cite{Amjad2018}
\cite{Sacristan2018} \cite{Sacristan2018b} \cite{CAMP2019} \cite{Convex2020} \cite{Lekidis2021} \cite{Poli2021}}  \\
& & Emergency electronic brake lights & \\
& & Queue/Traffic jam ahead warning  & \\
& & Road condition warning & \\ \cline{2-3} 
& Traffic Efficiency & Platooning (braking/join/leave) & \\
\hline
\multirow{2}{*}{V2V \textcolor{black}{bi-directional} communication} & Road Safety & See-through & \multirow{2}{3.5cm}{\cite{Perfecto2017} \cite{Thomas2018} \cite{Kutila2019} \cite{Convex2020}} \\  \cline{2-3}
& Convenience & Instant messaging & \\ 
\hline 
V2V multi-hop & Safety/Efficiency & Coverage extension of ITS messages & \cite{Khan2018} \cite{Wang2018b} \cite{Gupta2019b} \cite{Alghamdi2020} \\
\hline
\multirow{2}{*}{V2P periodic messages} & \multirow{2}{*}{Road Safety} &  Pedestrian collision warning to vehicles & 
\multirow{2}{*}{\cite{Emara2018} \cite{Nguyen2020}} \\
& & Vehicle collision warning to pedestrians & \\
\hline
\multirow{3}{*}{V2I periodic messages} & \multirow{2}{*}{Road Safety} & In-vehicle signage &  
\multirow{3}{3.5cm}{\cite{Xu2017} \cite{Khan2018} \cite{Guo2019} \cite{CAMP2019} \cite{Convex2020} \cite{Miao2021} }\\ 
& & Curve speed warning  &  \\ \cline{2-3}
& Traffic Efficiency & Green Light Optimized Speed Advisory & \\
\hline
\multirow{6}{*}{V2I event-driven messages} & \multirow{5}{*}{Road Safety} &  Infrastructure based collision warning & 
\multirow{6}{3.5cm}{\cite{5GAA2017} \cite{Napolitano2019} \cite{Nkenyereye2019} \cite{Kutila2019} \cite{Sequeira2019} \cite{Avino2019} \cite{Saxena2019} \cite{Giannone2020} \cite{Barmpounakis2020} \cite{Malinverno2020} 
\cite{Convex2020}  \cite{Tseng2020} \cite{Poli2021} \cite{Fu2021} \cite{Miao2021}} \\ 
& & Warning of vulnerable road user presence & \\
& & Infrastructure based traffic jam ahead warning & \\
& & Infrastructure based road condition warning & \\
& & Roadwork warning &  \\ \cline{2-3}
& Traffic Efficiency & Emergency vehicle signal preemption & \\
\hline
\multirow{4}{*}{RSU-relayed messages} & \multirow{4}{*}{Safety/Efficiency} & RSUs receive ITS messages and broadcast them to extend their reach & \multirow{4}{3.5cm}{\cite{Gupta2018} \cite{Gupta2019} \cite{Gupta2019b} \cite{Guo2019} \cite{Sequeira2019} \cite{Kang2019} \cite{Nkenyereye2019} \cite{Alghamdi2020}  \cite{Barmpounakis2020} \cite{Convex2020}}   \\
& & RSUs receive ITS messages and send them to a server (e.g., traffic control center) & \\
\hline
\multirow{7}{*}{V2N \textcolor{black}{bi-directional} communication} & Safety/Efficiency & Dynamic map download and update & 
\multirow{7}{3.5cm}{\cite{Xu2017} \cite{Wang2018b} \cite{Khan2018} \cite{Li2019} \cite{Kang2019} \cite{Ding2019} \cite{Chang2020} \cite{Convex2020}} \\ \cline{2-3}
& \multirow{6}{*}{Convenience} & Remote driving & \\ 
& & Media downloading & \\
& & Point of interest notification & \\
& & Payment services & \\
& & Stolen vehicle alert (warn service provider) & \\
& & Secure software updates for vehicles & \\ 
\hline
\end{tabular} 
\end{adjustbox}
\caption{Different types of V2X communication to support vehicular applications}
\label{table:commtypes}
\end{center}
\end{table}

\section{An exploration of the use of C-V2X communications to support connected vehicle applications}
\label{sec:exploration}

In this section we explore the contributions of recent literature regarding the use of C-V2X communications to support connected vehicle applications. Table~\ref{table:contributionsv1} situates the contributions of the different reviewed papers. The first row in the table represents the different types of communication used in connected vehicles applications that we have identified in section~\ref{sec:types}. The first column includes the different C-V2X related technologies reviewed in section~\ref{sec:technologies}, and also (separated by a double line) different alternatives to evaluate the performance or test the behavior of the technologies (simulations, field tests, and so on). Thus, the table allows to identify the types of V2X communication explored in each reviewed paper, the \mbox{C-V2X} technologies analyzed, and the validation methods employed.

\begin{table}[tb!]
\centering
\begin{adjustbox}{max width=0.94\textwidth}
\begin{tabular}{|c|p{3.5cm}|p{3.2cm}|c|c|c|c|p{3.2cm}|c|p{3.2cm}|}
\hline
 & \centering \rotatebox{90}{\textbf{V2V periodic messages}} & \centering \rotatebox{90}{\textbf{V2V event-driven messages}} & \centering \rotatebox{90}{\textbf{V2V \textcolor{black}{bi-directional} communication}} & \centering \rotatebox{90}{\textbf{V2V multi-hop}} & \centering \rotatebox{90}{\textbf{V2P periodic messages}} & \centering \rotatebox{90}{\textbf{V2I periodic messages}} & \centering \rotatebox{90}{\textbf{V2I event-driven messages}} & \centering \rotatebox{90}{\textbf{RSU-relayed messages}}  & \centering  \arraybackslash \rotatebox{90}{\textbf{V2N \textcolor{black}{bi-directional} communication}} \\ \hline
\multirow{3}{*}{\centering \textbf{IEEE 802.11p}} & 
\centering \vspace{-0.1cm} \cite{Vukadinovic2018} \cite{Bazzi2016} \cite{Bazzi2017} \cite{Calabuig2018} \cite{Molina-Masegosa2020} \cite{Xu2017} \cite{Shimizu2020} \cite{Shimizu2019} \cite{5GAA2019} \vspace{0.3cm} &
 &
 &
\multirow{3}{*}{\centering \cite{Alghamdi2020}} & 
 & 
\multirow{3}{*}{\centering \cite{Xu2017}} &
\centering \multirow{3}{3.1cm}{\centering  \cite{Xu2017} \cite{Saxena2019} \cite{Barmpounakis2020} \cite{Miao2021}} &
\multirow{3}{*}{\cite{Barmpounakis2020}} &
\centering \arraybackslash \multirow{3}{3.1cm}{\centering \cite{Xu2017}} 
\\
\hline
\multirow{3}{*}{\centering \textbf{LTE Uu}} & 
\centering \vspace{-0.1cm} \cite{Soleimani2017} \cite{Soleimani2017b} \cite{Peng2017} \cite{Peng2017b} \cite{Xu2017}  \cite{Kawasaki2017} \cite{Soleimani2018} \cite{Sacristan2018} \cite{Sacristan2018b} \cite{Mignardi2019} \vspace{0.3cm} & 
\centering \multirow{3}{3.1cm}{\centering \cite{Sacristan2018} \cite{Sacristan2018b}} &
 &
 &
\multirow{3}{*}{\cite{Nguyen2020} \cite{Emara2018}} &
\multirow{3}{*}{\cite{Khan2018} \cite{Xu2017} \cite{Miao2021}} &
\centering \multirow{3}{3.1cm}{\centering \cite{5GAA2017} \cite{Xu2017} \cite{Avino2019} \cite{Saxena2019} \cite{Napolitano2019} \cite{Giannone2020} \cite{Malinverno2020} \cite{Miao2021}} &
\multirow{3}{*}{\cite{Gupta2018} \cite{Gupta2019} \cite{Gupta2019b}} &
 \centering \arraybackslash \multirow{3}{3.1cm}{\centering \cite{Wang2018b} \cite{Khan2018} \cite{Xu2017} \cite{Convex2020}} 
\\
\hline
\multirow{2}{*}{\centering \textbf{5G NR Uu}} & \centering \multirow{2}{*}{\cite{Lekidis2021}} & 
\centering \multirow{2}{*}{\cite{Lekidis2021}} &
 \multirow{2}{*}{\centering \cite{Kutila2019}} &
 &
 &
 \multirow{2}{*}{\centering \cite{Miao2021}} &
\centering \vspace{-0.1cm} \cite{Kutila2019} \cite{Sequeira2019} \cite{Miao2021} \vspace{0.3cm} &
  &
\centering \arraybackslash \multirow{2}{3.1cm}{\centering \cite{Ding2019}} 
\\
\hline
\multirow{3}{*}{\textbf{eMBMS}} & 
 \centering \vspace{-0.1cm} \cite{Soleimani2018} \cite{Soleimani2017} \cite{Soleimani2017b} \cite{Peng2017} \cite{Peng2017b} \cite{Kawasaki2017} \cite{Sacristan2018} \cite{Sacristan2018b} \vspace{0.2cm} &
\centering \multirow{3}{3.1cm}{\centering \cite{Sacristan2018}\cite{Sacristan2018b}} &
 &
 &
 &
 &
 &
 &
\\
\hline
\multirow{2}{*}{\textbf{D2D}} & 
\centering \vspace{-0.1cm} \cite{Llatser2019} \cite{Gharba2017} \cite{Calabuig2018} \vspace{0.3cm} &
 &
 & \multirow{2}{*}{\centering \cite{Khan2018} \cite{Wang2018b}} &
 &
 &
 &
\multirow{2}{*}{\centering \cite{Alghamdi2020}} &
\\
\hline
\multirow{4}{*}{\textbf{Rel.~12 LTE PC5 mode 1}} & 
\centering \vspace{-0.1cm} \cite{Soleimani2018} \cite{Soleimani2017} \cite{Soleimani2017b} \cite{Gupta2018} \cite{Gupta2019} \cite{Sun2016} \cite{Peng2017} \cite{Peng2017b} \cite{Bazzi2016} \cite{Wang2018} \cite{Campolo2017} \vspace{0.3cm} &
\centering \multirow{4}{3.1cm}{\centering  \cite{Sun2016} \cite{Peng2017} \cite{Peng2017b} \cite{Wang2018}} &
 &
 &
 &
 &
 &
 &
\\
\hline
\multirow{2}{*}{\textbf{Rel.~12 LTE PC5 mode 2}} & 
\centering \vspace{-0.1cm} \cite{Gallo2017} \vspace{0.3cm} &
 &
 &
 &
 &
 &
 &
 &
\\
\hline
\multirow{3}{*}{\textbf{Rel.~14 LTE PC5 mode 3}} & 
\centering \vspace{-0.1cm} \cite{Vukadinovic2018} \cite{Peng2019} \cite{Geng2019} \cite{Bazzi2017} \cite{Kawasaki2017} \cite{Hegde2019} \cite{Sacristan2018b} \cite{Nardini2018} \vspace{0.3cm} &
 \centering \multirow{3}{3.1cm}{\centering \cite{Geng2019} \cite{Sacristan2018b}} & \multirow{3}{*}{\cite{Aslani2020} \cite{Guo2019b}} 
 &
\multirow{3}{*}{\centering \cite{Gupta2019b}} &
 &
\multirow{3}{*}{\centering \cite{Miao2021}} &
\centering \multirow{3}{3.1cm}{\centering \cite{Miao2021} \cite{Fu2021}} &
 &
\\
\hline
\multirow{4}{*}{\textbf{Rel.~14 LTE PC5 mode 4}} &
 \centering \vspace{-0.1cm} \cite{Vukadinovic2018} \cite{Park2018} \cite{Molina-Masegosa2020} \cite{Hirai2020} \cite{Molina-Masegosa2018} \cite{Schiegg2019} \cite{Sabeeh2019} \cite{Niebisch2020} \cite{Shimizu2020} \cite{Choi2020} \cite{Shimizu2019} \cite{CAMP2019} \cite{5GAA2019} \cite{Convex2020} \cite{Segata2021} \vspace{0.3cm} &
\centering \multirow{4}{3.1cm}{\centering \cite{CAMP2019} \cite{Convex2020} \cite{Poli2021}} &
\multirow{4}{*}{\centering \cite{Convex2020}} &
 &
 &
\multirow{4}{*}{\centering \cite{Miao2021} \cite{CAMP2019} \cite{Convex2020}} &
\centering \multirow{4}{3.1cm}{\centering \cite{Tseng2020} \cite{Poli2021} \cite{Miao2021}} &
 &
\\
\hline
\multirow{2}{*}{\textbf{Rel.~16 NR PC5 mode 1}} &  \centering \multirow{2}{*}{\cite{Hegde2021}}
 &
 &
 &
 &
 &
\multirow{2}{*}{\cite{Miao2021}} &
\centering \vspace{-0.1cm} \cite{Miao2021} \vspace{0.3cm} &
 &
\\
\hline
\multirow{2}{*}{\textbf{Rel.~16 NR PC5 mode 2}} & 
\centering \vspace{-0.1cm} \cite{Romeo2020} \cite{Romeo2020b} \vspace{0.3cm} &
\centering \multirow{2}{3.1cm}{\centering \cite{Romeo2020} \cite{Romeo2020b}} &
 &
 &
 &
\multirow{2}{*}{\centering \cite{Miao2021}} &
\centering \multirow{2}{3.1cm}{\centering \cite{Miao2021}} &
 &
\\
\hline
\multirow{2}{*}{\textbf{mmWave}} & 
 &
 &
\multirow{2}{*}{\cite{Perfecto2017}} &
 &
 &
 &
\centering \vspace{-0.1cm} \cite{Nkenyereye2019} \vspace{0.3cm} &
 \multirow{2}{*}{\centering \cite{Nkenyereye2019}} &
\centering \arraybackslash \multirow{2}{3.1cm}{\centering \cite{Li2019}} 
\\
\hline
\multirow{3}{*}{\textbf{Congestion control}} & 
\centering \vspace{-0.1cm} \cite{Calabuig2018} \cite{Hirai2020} \cite{Shimizu2020} \cite{Choi2020} \cite{Shimizu2019} \cite{CAMP2019} \vspace{0.3cm} &
\centering \multirow{3}{3.1cm}{\centering \cite{CAMP2019}} &
 &
 &
 &
\multirow{3}{*}{\centering \cite{CAMP2019}} &
 &
 &
\\
\hline
\multirow{2}{*}{\textbf{Cloud}} & 
 &
 &
 &
 &
 &
 &
\centering \vspace{-0.1cm} \cite{Barmpounakis2020} \cite{Napolitano2019} \vspace{0.3cm} &
\multirow{2}{*}{\centering \cite{Barmpounakis2020} \cite{Chang2020} } &
\centering \arraybackslash \multirow{2}{3.1cm}{\centering \cite{Convex2020}} 
\\
\hline
\multirow{3}{*}{\textbf{MEC/fog}} & 
\centering \multirow{3}{3.1cm}{\centering \cite{Sacristan2018} \cite{Sacristan2018b} \cite{Lekidis2021}} &
\centering \multirow{3}{3.1cm}{\centering  \cite{Sacristan2018}\cite{Sacristan2018b} \cite{Lekidis2021}} &
 &
 &
\multirow{3}{*}{\centering \cite{Nguyen2020} \cite{Emara2018}}&
&
\centering \vspace{-0.1cm} \cite{5GAA2017} \cite{Nkenyereye2019} \cite{Napolitano2019} \cite{Avino2019}  \cite{Giannone2020} \cite{Barmpounakis2020} \cite{Malinverno2020} \cite{Tseng2020} \cite{Poli2021} \vspace{0.3cm} &
\multirow{3}{*}{\centering \cite{Nkenyereye2019} \cite{Barmpounakis2020} \cite{Chang2020}} &
\centering \arraybackslash \multirow{3}{3.1cm}{\centering \cite{Ding2019} \cite{Convex2020}} 
\\
\hline
\multirow{2}{*}{\textbf{SDN}} & \centering \multirow{2}{*}{\cite{Lekidis2021}}
 & \centering \multirow{2}{*}{\cite{Lekidis2021}}
 &
 &
 &
 &
 &
 &
 &
\centering \arraybackslash \vspace{-0.1cm} \cite{Li2019} \vspace{0.3cm} 
\\
\hline
\multirow{2}{*}{\textbf{Network virtualization}} & \centering \multirow{2}{*}{\cite{Lekidis2021}}
 & \centering \multirow{2}{*}{\cite{Lekidis2021}}
 &
 &
 &
 &
 &
\centering \vspace{-0.1cm} \cite{Barmpounakis2020} \cite{Avino2019} \vspace{0.3cm} &
\multirow{2}{*}{\centering \cite{Barmpounakis2020}} &
\centering \arraybackslash \multirow{2}{3.1cm}{\centering \cite{Convex2020}} 
\\
\hline
\multirow{2}{*}{\textbf{Network slicing}} & \centering \multirow{2}{*}{\cite{Lekidis2021}}
 & \centering \multirow{2}{*}{\cite{Lekidis2021}}
 &
 &
 &
 &
\multirow{2}{*}{\centering \cite{Khan2018}} &
\centering \multirow{2}{*}{\cite{5GAA2017}}  &
 &
 \centering \arraybackslash \vspace{-0.1cm} \cite{Khan2018} \cite{Ding2019} \cite{Convex2020} \vspace{0.3cm}
\\
\hline
\hline
\multirow{5}{*}{\textbf{\textcolor{black}{Focused simulations}}} & 
\centering \vspace{-0.1cm}   \cite{Sun2016} \cite{Bazzi2016} \cite{Bazzi2017} \cite{Peng2017} \cite{Peng2017b} \cite{Gallo2017}  \cite{Kawasaki2017} \cite{Gharba2017} \cite{Campolo2017} \cite{Wang2018} \cite{Geng2019} \cite{Peng2019} \cite{Shimizu2019} \cite{Mignardi2019} \cite{Schiegg2019} \cite{Romeo2020} \cite{Romeo2020b} \cite{Choi2020}  \vspace{0.3cm} &
\centering \multirow{5}{3.1cm}{\centering \cite{Sun2016} \cite{Peng2017} \cite{Peng2017b} \cite{Geng2019} \cite{Wang2018} \cite{Romeo2020} \cite{Romeo2020b}} &
\multirow{5}{*}{\centering \cite{Perfecto2017} \cite{Aslani2020} \cite{Guo2019b}} &
\multirow{5}{*}{\centering  \cite{Wang2018b}} &
\multirow{5}{*}{\centering \cite{Emara2018}}  &
  &
\multirow{5}{3.1cm}{\centering \cite{Nkenyereye2019} \cite{Sequeira2019} \cite{Saxena2019} \cite{Tseng2020}} &
\multirow{5}{*}{\centering \cite{Nkenyereye2019}} &
\centering \arraybackslash \multirow{5}{3.1cm}{\centering \cite{Wang2018b} \cite{Ding2019}} 
\\
\hline
\multirow{5}{*}{\textbf{System-level simulations}} & 
\centering \vspace{-0.1cm} \cite{Vukadinovic2018} \cite{Soleimani2018} \cite{Soleimani2017} \cite{Soleimani2017b} \cite{Gupta2018} \cite{Gupta2019} \cite{Park2018} \cite{Sacristan2018} \cite{Hegde2019} \cite{Calabuig2018} \cite{Sacristan2018b} \cite{Molina-Masegosa2020} \cite{Hirai2020} \cite{Nardini2018} \cite{Molina-Masegosa2018} \cite{Sabeeh2019} \cite{Shimizu2020} \cite{Segata2021} \cite{Hegde2021} \vspace{0.3cm}  &
\centering \multirow{5}{3.1cm}{\centering \cite{Sacristan2018} \cite{Sacristan2018b} \cite{Poli2021}} &
 &
\multirow{5}{*}{\centering \cite{Khan2018} \cite{Gupta2019b} \cite{Alghamdi2020}} &
\multirow{5}{*}{\centering  \cite{Nguyen2020}} &
\multirow{5}{*}{\centering  \cite{Khan2018}} & \centering \multirow{5}{*}{\cite{Malinverno2020} \cite{Poli2021} \cite{Fu2021}} &
\multirow{5}{*}{\centering \cite{Gupta2018} \cite{Gupta2019} \cite{Gupta2019b} \cite{Alghamdi2020}} &
\centering \arraybackslash \multirow{5}{3.1cm}{\centering \cite{Khan2018}} 
\\
\hline
\multirow{2}{*}{\textbf{Real equipment tests}} & 
\centering \vspace{-0.1cm} \cite{5GAA2019} \vspace{0.3cm} &
 &
 &
 &
 &
 &
\centering \multirow{2}{3.1cm}{\centering \cite{Giannone2020} \cite{Avino2019}} &
  &
\\
\hline
\multirow{3}{*}{\textbf{Field tests}} & 
\centering \vspace{-0.1cm} \cite{Gharba2017} \cite{Xu2017} \cite{Niebisch2020} \cite{CAMP2019} \cite{5GAA2019} \cite{Convex2020} \cite{Lekidis2021} \vspace{0.3cm} &
\centering \multirow{3}{3.1cm}{\centering \cite{CAMP2019} \cite{Convex2020} \cite{Lekidis2021}} &
\multirow{3}{*}{\centering \cite{Kutila2019} \cite{Convex2020}} &
 &
 &
\multirow{3}{*}{\centering \cite{Xu2017} \cite{Miao2021} \cite{CAMP2019} \cite{Convex2020}} &
 \centering \multirow{3}{3.1cm}{\centering \cite{5GAA2017} \cite{Xu2017} \cite{Napolitano2019} \cite{Barmpounakis2020}  \cite{Miao2021}} &
\multirow{3}{*}{\centering \cite{Barmpounakis2020}} &
 \centering \arraybackslash \multirow{3}{3.3cm}{\centering \cite{Li2019} \cite{Xu2017} \cite{Convex2020}} 
\\
\hline
\multirow{2}{*}{\textbf{Demonstrations}} & 
\centering \vspace{-0.1cm} \cite{Llatser2019} \cite{Convex2020} \vspace{0.3cm} &
\centering \multirow{2}{3.1cm}{\centering \cite{Convex2020}} &
 \multirow{2}{*}{\centering  \cite{Convex2020}} &
 &
 &
\multirow{2}{*}{\centering \cite{Miao2021} \cite{Convex2020}} &
\centering \multirow{2}{3.1cm}{\centering \cite{Miao2021}} &
 &
\centering \arraybackslash \multirow{2}{3.1cm}{\centering \cite{Convex2020}}
\\
\hline
\end{tabular} 
\end{adjustbox}
\caption{Literature on the use of C-V2X communications to support connected vehicle applications}
\label{table:contributionsv1}
\end{table}

The list of \mbox{C-V2X} related technologies includes \textbf{D2D}, by which we mean papers that assume generic direct-to-direct communication technology without following a particular 3GPP solution, or papers in which it is not clear which D2D solution is used. \textit{IEEE 802.11p} is also on the list even though it is not a C-V2X technology as it is often used in performance comparisons with \mbox{C-V2X} technologies, or in hybrid systems combining \textit{IEEE 802.11p} with C-V2X technologies.

With respect to the validation part, we have the following categories:

\begin{enumerate}
\item \textbf{\textcolor{black}{Focused simulations}:}  this refers to numerical analysis or simulations that do not model the whole system under study (for example, they model radio issues but not the MAC protocol, or they focus on the performance of particular algorithms within the system). We also report here papers that do not provide enough details about what exactly was simulated or how.
\item \textbf{System-level simulations:} simulations that model the whole system under study and are used to experimentally analyze the system. 
\item \textbf{Real equipment tests:} real-life equipment in a lab environment is used to experimentally analyze the system under study.
\item \textbf{Field tests:} tests with real equipment are performed in open roads, or in test tracks that emulate real conditions. 
\item \textbf{Demonstrations:} complete applications are shown in open roads for demonstration purposes. The aim is not to have an environment in which to carry out test and obtain performance numbers, but show the technology in real life to stakeholders. 
\end{enumerate}     
 
\textcolor{black}{Table~\ref{table:contributionsv1} shows that the interest on communication types is technology-agnostic, since most works are focused on V2V periodic messages and V2I event-driven ones. Regarding technologies, the more popular ones are LTE Uu (due to its wide availability), and the sidelink communications modes. Although in LTE Rel. (Release) 12 almost all works were focused on the network-coordinated mode 1, on LTE Rel.~14 (and in lesser extent NR Rel.~16, due to the few papers considering it) the research interest seems to be moving to the distributed-coordination modes. Besides the obvious technological evolution (with older technologies accumulating more works), the validation part of Table~\ref{table:contributionsv1} shows that a great majority of papers are still simulation-based, with a small number of works using real-world equipment, which in many cases are just using an LTE Uu interface (specially on V2I applications), albeit the availability of Rel.~14 LTE equipment is enabling several works with a real-world evaluation.}
 
In the rest of the section, we provide more detailed explanations about the contributions of the reviewed papers to the analysis of using C-V2X to support applications for the connected vehicle. The structure follows the classification of different types of V2X communication introduced in section~\ref{sec:types}.

\subsection{V2V periodic messages}

This is the most common communication type to support applications for the connected vehicle. As such, the use of C-V2X technologies to provide this communication type has been the most studied in the literature. 

Many of the early efforts in the application of cellular communication technologies to support connected vehicle applications consisted in the study of using the Proximity Services (ProSe) D2D interface, defined in Rel.~12, in vehicular scenarios (Rel.~12 LTE PC5 mode 1 \cite{Sun2016} and mode 2 \cite{Gallo2017}). Since this D2D interface was not designed with vehicular communication requirements in mind, the focus was on identifying its limitations for supporting V2V periodic messages, and proposing alternative resource allocation strategies with the aim of maximizing the use of the medium while meeting the latency and reliability constraints of V2V communications for safety applications.  

Most early works used simulations in the evaluation part but, probably due to the lack of available tools, these simulations did not model the whole system, but only parts of it (such as algorithms for resource scheduling), or included significant simplifications of parts of the system. Even so, the importance of tests in the real world was already considered from the beginning: the work in \cite{Gharba2017} used software defined radios and two real vehicles to get a first look at the direct communication between vehicles using cellular-like radios, while \cite{Xu2017} compared the use of IEEE 802.11p for direct V2V communications with the use of communications through the network in the infrastructure using LTE Uu (in the latter case, the latency was shown to be too large for safety applications). 

Several initial works compared the performance of IEEE 802.11p and Rel.~12 LTE PC5 mode 1 (resources assigned from the infrastructure) for sending V2V periodic messages, showing potential benefits of the LTE PC5 approach by applying strategies of frequency reuse \cite{Bazzi2016} \cite{Bazzi2017} \cite{Calabuig2018}. 

There was also interest in comparing the performance of sending the periodic messages from a source vehicle to the infrastructure (LTE Uu) to be later broadcast from there to target vehicles with the use of D2D approaches (Rel.~12 LTE PC5 mode 1 \cite{Soleimani2017} \cite{Soleimani2017b} \cite{Soleimani2018} and  Rel.~14 LTE PC5 mode 3 \cite{Kawasaki2017}), which concluded that latency was lower with the D2D approaches. The work in \cite{Sacristan2018} (extended in \cite{Sacristan2018b}) used servers deployed at base stations to improve the latency of the communications of the V2V periodic messages sent through the infrastructure (a MEC-like approach), and concluded that the performance from the point of view of latency could be enough to support safety services, but performance degrades with network load. 

Several papers \cite{Soleimani2017} \cite{Soleimani2017b} \cite{Soleimani2018} \cite{Sacristan2018b} \cite{Gupta2018} \cite{Gupta2019} explored the combination of D2D and \mbox{LTE Uu}, by using information obtained from the messages sent by vehicles and received in the infrastructure to improve the scheduling of resources in the network for D2D communications. For example, \cite{Sacristan2018b} \cite{Gupta2018} \cite{Gupta2019} proposed to reuse frequencies for D2D communications according to vehicles' locations obtained from the periodic messages they send to the infrastructure.  

\textcolor{black}{Another work that explored the use of the LTE Uu interface for sending V2V periodic messages was \cite{Mignardi2019}. The studied scenario involved mobile base stations deployed in Unmanned Aerial Vehicles (UAV) to improve the connectivity service offered to vehicles compared with using only terrestrial base stations. The analyzed application was collective perception (extended sensors), which involves broadcasting messages with sensor data (through the infrastructure in this case) to increase the sensing range beyond that of the sensors of individual vehicles. The main conclusion of this work is identifying the potential of UAVs to improve connectivity in vehicles.} 

Rel.~14 of 3GPP standards included the first D2D solution specifically developed for communications among vehicles. Different works in the literature have studied its performance when sending V2V periodic messages and proposed improvements to better fulfill the communication requirements of the connected vehicle. Rel.~14 LTE PC5 mode 4 was studied in \cite{Park2018} \cite{Molina-Masegosa2018} \cite{Sabeeh2019} \cite{Schiegg2019} \cite{Hirai2020} \cite{Choi2020}. 

The work in \cite{Park2018} proposed a feedback mechanism to deal with one of the limitations of the distributed scheduling algorithm of the standard, namely the possibility of repeated collisions among vehicles choosing the same resource. In \cite{Molina-Masegosa2018}, authors explored different values for the configuration parameters of Rel.~14 LTE PC5 mode 4 interfaces to achieve good performance when sending V2V periodic messages. The performance was measured in terms of packet delivery ratio and inter-packet gaps, and results showed a big impact on performance stemming from different configuration options, and the poor performance it exhibits in high network traffic load situations. In \cite{Sabeeh2019}, the performance of the S-SPS algorithm specified in the standard for resource allocation was evaluated; identifying weaknesses such as persistent half-duplex errors, hidden terminal errors, and persistent packet collision errors, which result in poor performance in situations of high network traffic load; and also an alternative algorithm for resource allocation was proposed and evaluated. \textcolor{black}{The work in \cite{Schiegg2019} developed an analytical model to study the performance of a collective perception sharing application based on Rel.~14 LTE PC5 mode 4. Collective perception was studied with simulations in \cite{Schiegg2021} but, although Rel.~14 LTE PC5 mode 4 is considered as potential technology for the implementation of the solution, the simulations were done with IEEE 802.11p}.  In \cite{Hirai2020}, the authors studied the applicability of Rel.~14 LTE PC5 mode 4 to a particular application: a crash warning system in scenarios with network congestion, and concluded that improvements are needed to this technology in order to support this application in high density scenarios. In \cite{Choi2020}, a novel congestion control mechanism for Rel.~14 LTE PC5 mode 4 was proposed, which is able to support two different communication service classes (better coverage or better rate).  

The comparison of C-V2X with IEEE 802.11p for sending V2V periodic messages has been revisited, now comparing vehicular networks based on IEEE~802.11p (DSRC or ITS-G5) with its most straightforward alternative, the Rel.~14 LTE PC5 mode 4 (direct communications without support from the infrastructure): In \cite{Molina-Masegosa2020} and \cite{Shimizu2019}, IEEE~802.11p and Rel.~14 LTE PC5 mode 4 were compared considering realistic generation of CAMs \cite{ETSI2019}, according to the dynamics of vehicles and with changes in the sizes of the messages, and they found that the performance (packet reception ratio and inter-packet gap) of IEEE
~802.11p is better in scenarios with medium and high vehicle traffic (also when using congestion control \cite{Shimizu2019}) because the S-SPS algorithm, defined in LTE PC5 mode 4, is found to have limitations when transmitting messages at shifting periods and with variable sizes. IEEE~802.11p also achieves less latency, due to the long time a vehicle has to wait in LTE PC5 mode 4 for the persistent scheduled resource. As in previous works, the persistent collision and half-duplex problems were identified as sources of performance loss in Rel.~14 LTE PC5 mode 4. In \cite{Shimizu2020}, BSMs (with fixed size and frequency) were sent instead of CAMs, and SAE~J2945/1 was used as congestion control algorithm in both technologies, achieving similar performance at low distances (until up to 300m) if HARQ was used in Rel.~14 LTE PC5 mode 4 (which means double bandwidth is needed), while the performance of Rel.~14 LTE PC5 mode 4 was better at longer distances.  The consecutive packet loss issue of Rel.~14 LTE PC5 mode 4, due to the channel access mechanisms, was also confirmed in this study. These works \cite{Molina-Masegosa2020} \cite{Shimizu2019} \cite{Shimizu2020} used different system-level simulator tools in the validation part.

 In \cite{5GAA2019}, the comparison between IEEE~802.11p and Rel.~14 LTE PC5 mode~4 for sending V2V periodic messages was performed using real equipment by means of lab tests and experiments on a test track. Results showed that radio coverage and robustness against interference is better for Rel.~14 LTE PC5 mode 4, which achieves longer communication ranges and better performance in the presence of obstacles. The experiments also tested the SAE J2945/1 congestion control algorithm, originally developed for vehicular networks based on IEEE 802.11p, with the Rel.~14 LTE PC5 mode 4 network, and showed that it works well avoiding large degradation of performance with increased network load. Experiments in test tracks and open roads using real Rel.~14 LTE PC5 mode 4 equipment for sending V2V periodic messages were also reported in \cite{CAMP2019}, \cite{Niebisch2020}, and \cite{Convex2020}. These works explored the performance of using different configuration options (number of antennas, use of HARQ, transmit power, modulation and coding scheme, bandwidth, congestion control), and in different situations (line-of-sight/non-line of sight, moving/static vehicles, up to eight vehicles in a test track and in open roads). Interestingly, \cite{Niebisch2020} used a realistic antenna placement in the vehicle (encasing and location in the car similar to which could be done in a real vehicle) and this has a significant impact on performance (packet error rate as high as 16.65\% at distances lower than 100m while, for example \cite{Convex2020} achieved good performance (minimum losses) at even 1.2~km ranges. In \cite{Convex2020}, several applications were implemented and demonstrated, so not only the communication part was analyzed, but also elements such as user interfaces and information processing. Examples of tested applications based on periodic CAMs were blind spot warning/lane change warning, intersection moving assist, left turn assist, and follow-me information. No issues were found with the operation of any of these applications.  
 
\textcolor{black}{We end the section with two tables summarizing some key points. Table~\ref{table:researchtopics_V2Vperiodic} classifies the main contributions of the surveyed articles on the use of V2V periodic messages, and Table~\ref{table:applications_V2Vperiodic} lists specific example applications analyzed in those articles (note that many articles study the exchange of messages for awareness, for example, but without particularizing to specific applications such as collision warning or cooperative adaptive cruise control).}

\begin{table}[tb!] \footnotesize
\begin{center}
\begin{adjustbox}{max width=1.1\textwidth}
\begin{tabular}{|l|p{2.3cm}|}
\hline
\centering \textbf{Contributions} & \textbf{Related papers} \\ \hline
Network-based resource allocation in PC5 interface & \centering \arraybackslash \cite{Sun2016} \cite{Bazzi2016} \cite{Bazzi2017} \cite{Calabuig2018} \\ \hline
\multirow{2}{*}{\centering Distributed resource allocation in PC5 interface} & \centering \arraybackslash \cite{Gallo2017} \cite{Park2018} \cite{Sacristan2018b} \cite{Sabeeh2019} \cite{Molina-Masegosa2020} \cite{Shimizu2019} \cite{Shimizu2020} \\ \hline
\multirow{2}{*}{\centering Parameter configuration of Rel.~14 LTE PC5 mode 4} & \centering \arraybackslash \cite{Molina-Masegosa2018} \cite{Shimizu2020} \cite{CAMP2019}  \cite{Niebisch2020} \cite{Convex2020} \\ \hline
\multirow{3}{*}{\centering Comparison of IEEE 802.11p with C-V2X} &  \centering \arraybackslash \cite{Xu2017} \cite{Bazzi2016} \cite{Bazzi2017} \cite{Calabuig2018}  \cite{Soleimani2017} \cite{Soleimani2017b} \cite{Soleimani2018} \cite{Kawasaki2017} \cite{Molina-Masegosa2020} \cite{Shimizu2019} \cite{Shimizu2020} \cite{5GAA2019} \\ \hline
\multirow{2}{*}{\centering V2V communications through the Uu interface} & \centering \arraybackslash \cite{Xu2017} \cite{Soleimani2017} \cite{Soleimani2017b} \cite{Soleimani2018} \cite{Kawasaki2017} \cite{Sacristan2018} \cite{Sacristan2018b} \cite{Mignardi2019} \\ \hline
\multirow{2}{*}{\centering Combining Uu and D2D for providing V2V services} & \centering \arraybackslash \cite{Soleimani2017} \cite{Soleimani2017b} \cite{Soleimani2018} \cite{Sacristan2018b} \cite{Gupta2018} \cite{Gupta2019} \\ \hline
Edge processing strategies & \centering \arraybackslash \cite{Sacristan2018} \cite{Sacristan2018b} \\ \hline
Performance impact of network load & \centering \arraybackslash \cite{Sacristan2018} \cite{Sacristan2018b}  \cite{Schiegg2019} \cite{Hirai2020} \\ \hline
Congestion control & \centering \arraybackslash \cite{Choi2020} \cite{Shimizu2019} \cite{Shimizu2020} \cite{5GAA2019} \\ \hline 
Demonstration/analysis of specific applications &  \centering \arraybackslash \cite{Schiegg2019} \cite{Mignardi2019} \cite{Hirai2020} \cite{Convex2020} \\ \hline
\multirow{2}{*}{\centering Performance evaluation tested with real-life equipment} & \centering \arraybackslash \cite{Gharba2017} \cite{5GAA2019} \cite{CAMP2019}  \cite{Niebisch2020} \cite{Convex2020} \\ \hline
\end{tabular} 
\end{adjustbox}
\caption{\textcolor{black}{Classification of contributions of surveyed papers, V2V periodic messages}}
\label{table:researchtopics_V2Vperiodic}
\end{center}
\end{table}

\begin{table}[tb!] \footnotesize
\begin{center}
\begin{adjustbox}{max width=1.1\textwidth}
\begin{tabular}{|l|p{2.3cm}|}
\hline
\centering \textbf{Specific applications} & \textbf{Related papers} \\ \hline
Collective perception & \centering \arraybackslash \cite{Mignardi2019} \cite{Schiegg2019} \\ \hline
Collision warning & \centering \arraybackslash \cite{Hirai2020} \\ \hline
Blind spot warning/lane change warning & \centering \arraybackslash \cite{Convex2020} \\ \hline
Intersection moving assist & \centering \arraybackslash \cite{Convex2020} \\ \hline
Left turn assist & \centering \arraybackslash \cite{Convex2020} \\ \hline
Follow-me information & \centering \arraybackslash \cite{Convex2020} \\ \hline
\end{tabular} 
\end{adjustbox}
\caption{\textcolor{black}{Examples of specific applications analyzed on surveyed papers, V2V periodic messages}}
\label{table:applications_V2Vperiodic}
\end{center}
\end{table}

\subsection{V2V event-driven messages}
\label{sec:v2vevent}

An early work \cite{Sun2016} considered sending messages within a deadline using the Rel.~12 LTE PC5 mode 1 technology, and proposing an algorithm for resource allocation. 

Besides V2V periodic messages, the work in \cite{Sacristan2018} (extended in \cite{Sacristan2018b}) also tested sending event-driven V2V messages from a vehicle to the infrastructure, to be later broadcast from there or transmitted in unicast to some particular vehicles. The event-driven messages alerted vehicles of dangerous situations (e.g., collision avoidance). This work showed that using servers deployed in the base stations (a MEC-like approach) allows to achieve a performance that was considered good enough for safety services, but the performance degrades with network load, even for the event-driven messages that had more priority than the periodic messages (CAMs) being sent. In \cite{Sacristan2018b}, this approach was compared with sending the messages directly between vehicles using the Rel.~14 LTE PC5 mode 3 technology, and the performance was better in the latter case (although just single-cell scenarios were considered).  

In \cite{Poli2021}, the LTE PC5 mode 4 technology was used to implement a cooperative lane merging application based on sending V2V event-driven messages combined with periodic awareness messages.  The success ratio achieved by the application was larger than 90\% if the delay in the operation of the application was very low (less than 5ms), but degrades with larger delays (delays that can be originated in the communications part or in the application information processing).     

The main focus in the design of Rel.~14 LTE PC5 interfaces was on supporting V2V periodic messages. So, the limitations of Rel.~14 LTE PC5 mode 4 interfaces to send V2V event-driven messages were identified in \cite{Romeo2020} \cite{Romeo2020b}. In particular, limitations were found in the autonomous sidelink resource allocation algorithm (S-SPS). For example, the sensing window of 1s used in Rel.~14 LTE PC5 mode 4 works well with periodic messages but it is too long for asynchronous traffic. The work in \cite{Romeo2020} \cite{Romeo2020b} explored several strategies and configuration parameters, following the work in progress at the time in Rel.~16 NR PC5 mode 2 \cite{3GPP2020b}, for the resource allocation algorithm for event-driven messages. The main allocation options explored were: random resource selection, sensing-based resource selection (S-SPS as in Rel.~14 LTE PC5 mode 4), and short-term sensing (S-SPS as in Rel.~14 LTE PC5 mode 4, but with a 100ms sensing window instead of 1s). These allocation options were studied while the event-driven traffic co-exists with periodic messages with resources allocated using the S-SPS of Rel.~14 LTE PC5 mode 4. The conclusion was that sensing is important for performance, but that short-term sensing (100ms) is better for event-driven messages. The experimental part of the work was done using system-level simulations.

Experiments with V2V event-driven messages using real Rel.~14 LTE PC5 mode 4 equipment in test tracks and open roads were reported in \cite{CAMP2019} and \cite{Convex2020}. These works explored the performance of different configuration options (number of antennas, use of HARQ, transmit power, modulation and coding scheme, bandwidth, congestion control), and in different situations (line-of-sight/non-line of sight, moving/static vehicles, up to eight vehicles in a test track and in open roads). As mentioned in previous section, \cite{Convex2020} also reported the implementation and demonstration of several applications. Some of them were based on sending event-driven messages, including emergency electronic brake light and slow stationary vehicle warning. These demonstrations involved two vehicles and a motorcycle operating under different conditions and speeds, and no issues were found with the operation of the applications.   

\textcolor{black}{We end the section with two tables summarizing some key points. Table~\ref{table:researchtopics_V2Veventdriven} classifies the main contributions of the surveyed articles that study the use of V2V event-driven messages, and Table~\ref{table:applications_V2Veventdriven} lists specific example applications analyzed in those articles.}

\begin{table}[tb!] \footnotesize
\begin{center}
\begin{adjustbox}{max width=1.1\textwidth}
\begin{tabular}{|l|p{2.3cm}|}
\hline
\centering \textbf{Contributions} & \textbf{Related papers} \\ \hline
Network-based resource allocation in PC5 interface & \centering \arraybackslash \cite{Sun2016} \cite{Sacristan2018b}\\ \hline
Distributed resource allocation in PC5 interface & \centering \arraybackslash \cite{Poli2021} \cite{Romeo2020} \cite{Romeo2020b} \\ \hline
\centering Parameter configuration of PC5 interface & \centering \arraybackslash  \cite{Romeo2020} \cite{Romeo2020b} \cite{CAMP2019} \cite{Convex2020} \\ \hline
V2V communications through the Uu interface & \centering \arraybackslash  \cite{Sacristan2018} \cite{Sacristan2018b} \\ \hline
Edge processing strategies & \centering \arraybackslash \cite{Sacristan2018} \cite{Sacristan2018b} \\ \hline
Performance impact of network load & \centering \arraybackslash  \cite{Sacristan2018} \cite{Sacristan2018b} \\ \hline
Demonstration/analysis of specific applications & \centering \arraybackslash \cite{Convex2020} \cite{Poli2021} \\ \hline
Performance evaluation tested with real-life equipment & \centering \arraybackslash \cite{CAMP2019} \cite{Convex2020} \\ \hline
\end{tabular} 
\end{adjustbox}
\caption{\textcolor{black}{Classification of contributions of surveyed papers, V2V event-driven messages}}
\label{table:researchtopics_V2Veventdriven}
\end{center}
\end{table}

\begin{table}[tb!] \footnotesize
\begin{center}
\begin{adjustbox}{max width=1.1\textwidth}
\begin{tabular}{|l|p{2.3cm}|}
\hline
\centering \textbf{Specific applications} & \textbf{Related papers} \\ \hline
Cooperative lane merging & \centering \arraybackslash \cite{Poli2021} \\ \hline
Emergency electronic brake light & \centering \arraybackslash \cite{Convex2020} \\ \hline
Slow stationary vehicle warning & \centering \arraybackslash \cite{Convex2020} \\ \hline
\end{tabular} 
\end{adjustbox}
\caption{\textcolor{black}{Examples of specific applications analyzed on surveyed papers, V2V event-driven messages}}
\label{table:applications_V2Veventdriven}
\end{center}
\end{table}

\subsection{V2V periodic messages and V2V event-driven messages: Platooning}

An application that has received particular attention is platooning, i.e., the support of several vehicles moving together at short inter-vehicle distances and following a platoon leader. Platooning improves the efficiency of road use and fuel efficiency in vehicles. Platooning is implemented with V2V periodic messages, used by vehicles to exchange the information they need to operate safely within the platoon; and V2V event-driven messages, used by vehicles to join and leave the platoon or to warn of an emergency braking situation. Because of its interest, and the combination of V2V periodic and event-driven messages, we dedicate this section to analyze the research on using C-V2X technologies to support platooning applications. 

Different works have explored the communications needed to support a platoon and how to implement them with C-V2X technologies. A common model is to have V2V periodic messages between the platoon leader and its members, and from each member to the follower vehicle. Initial works explored the use of Rel.~12 LTE PC5 mode 1 technology \cite{Campolo2017} \cite{Peng2017} \cite{Peng2017b} \cite{Wang2018} \cite{Geng2019} to support these communications, studying algorithms for resource allocation to maximize the reuse of radio resources and to minimize latency. In some cases, the communication between the platoon leader and its members is implemented by sending the information to the infrastructure from the platoon leader, and multicast from there to the platoon members using eMBMS \cite{Peng2017} \cite{Peng2017b}. The communications needed for platoon formation and, in some cases, braking events, were also considered in \cite{Peng2017} \cite{Peng2017b} \cite{Wang2018} \cite{Geng2019}. These works used simulations in the evaluation part, but they did not model the whole system, only parts of it (such as algorithms for resource scheduling), or included significant simplifications in parts of the system.

Later works already considered the use of D2D technologies specifically designed for \textcolor{black}{vehicular communications} to support the platooning application: Rel.~14 LTE PC5 mode 3 \cite{Nardini2018} \cite{Vukadinovic2018} \cite{Peng2019} \cite{Hegde2019}; Rel.~14 LTE PC5 mode 4 \cite{Vukadinovic2018} \cite{Segata2021}; and Rel.~16 5G NR mode 1 \cite{Hegde2021}. In the works using Rel.~14 LTE PC5 mode 3 and Rel.~16 5G NR mode 1, the focus was on exploring different algorithms for resource allocation. The works in \cite{Hegde2019} and \cite{Hegde2021} considered scenarios with several cells, which is an important issue when resources are allocated from the network. In both works, a group scheduling approach to resource assignment was proposed in which resources are assigned to the platoon as a whole through the platoon leader, and not individually to each vehicle. This mechanism improves performance during handovers but requires coordination among base stations. Even with group scheduling, at high network load scenarios the works concluded that further improvements are needed to meet platoon application requirements. In \cite{Vukadinovic2018}, a comparison between IEEE 802.11p, Rel.~14 LTE PC5 mode 3, and Rel.~14 LTE PC5 mode 4 was done, and the conclusion was that PC5 modes can achieve reduced inter-truck spacing due to more reliable communications, and that Rel.~14 LTE PC5 mode 3 outperforms Rel.~14 LTE PC5 mode 4, but note that mode 3 was studied considering only scenarios with just one cell. The work in \cite{Segata2021} explored the limitations of the S-SPS algorithm for resource scheduling in LTE PC5 mode 4, in particular the half-duplex problem and the collision bursts, and the conclusion was that, although the global performance (average packet delivery ratio) seems adequate, the concentration of losses have a significant impact on the performance of the platooning application. The evaluation of all these later works was based on system-level simulations using different simulation tools. 

In \cite{CAMP2019}, the performance of communications in a platoon, using real Rel.~14 LTE PC5 mode 4 equipment and in the presence of network congestion, was tested in a test track. The main conclusion was the importance of using a congestion control mechanism since performance, measured as packet error rate or information age, degrades sharply without it. However, no specific measures of the inter-vehicle distance supported in the tested scenarios were given.  

The work in \cite{Lekidis2021} proposed a framework that combines NFV, SDN, MEC, and network slicing to deploy ultra reliable and low latency vehicular services. The framework was deployed in a test track using 5G NR Uu technology to test a platooning application. The achieved latency for periodic CAMs and event-driven messages, thanks to using a specific slice for low latency and deploying a 5G core as a virtual function in the edge, was below 10ms, which fulfils the requirement set by 3GPP for platooning applications. \textcolor{black}{Table~\ref{table:researchtopics_platooning} classifies the main contributions of the surveyed articles that study the use of V2V periodic messages and V2V event-driven messages in platooning applications.}

\begin{table}[tb!] \footnotesize
\begin{center}
\begin{adjustbox}{max width=1.1\textwidth}
\begin{tabular}{|l|p{2.3cm}|}
\hline
\centering \textbf{Contributions} & \textbf{Related papers} \\ \hline
\multirow{2}{*}{Communication models to support platooning} & \centering \arraybackslash \cite{Campolo2017} \cite{Peng2017} \cite{Peng2017b} \cite{Wang2018} \cite{Geng2019} \\ \hline
\multirow{3}{*}{Network-based resource allocation in PC5 interface} & \centering \arraybackslash \cite{Campolo2017} \cite{Peng2017} \cite{Peng2017b} \cite{Wang2018} \cite{Geng2019} \cite{Nardini2018} \cite{Vukadinovic2018} \cite{Peng2019} \cite{Hegde2019} \cite{Hegde2021} \\ \hline
Distributed resource allocation in PC5 interface & \centering \arraybackslash  \cite{Vukadinovic2018} \cite{Segata2021} \\ \hline
Comparison of IEEE 802.11p with C-V2X & \centering \arraybackslash \cite{Vukadinovic2018} \\ \hline 
V2V communications through the Uu interface & \centering \arraybackslash \cite{Peng2017} \cite{Peng2017b} \cite{Lekidis2021} \\ \hline
Edge processing strategies & \centering \arraybackslash \cite{Lekidis2021} \\ \hline
Performance impact of network load & \centering \arraybackslash \cite{Hegde2019} \cite{Hegde2021} \cite{CAMP2019} \\ \hline
\multirow{4}{*}{Demonstration/analysis of specific applications} & \centering \arraybackslash \cite{Campolo2017} \cite{Peng2017} \cite{Peng2017b} \cite{Wang2018} \cite{Geng2019} \cite{Nardini2018} \cite{Vukadinovic2018} \cite{Peng2019} \cite{Hegde2019} \cite{Hegde2021} \cite{Segata2021} \cite{CAMP2019} \cite{Lekidis2021} \\ \hline
Performance evaluation tested with real-life equipment & \centering \arraybackslash \cite{CAMP2019} \cite{Lekidis2021} \\ \hline
\end{tabular} 
\end{adjustbox}
\caption{\textcolor{black}{Classification of contributions of surveyed papers, platooning}}
\label{table:researchtopics_platooning}
\end{center}
\end{table}

\subsection{V2V \textcolor{black}{bi-directional} communication}

There are not many works exploring V2V \textcolor{black}{bi-directional} communications because D2D technologies developed by 3GPP, at least previously to Rel.~16 NR PC5, are not designed for point-to-point \textcolor{black}{bi-directional} communications in vehicular networks. Even so, there are relevant scenarios that require this type of communication, in particular, see-through applications: a vehicle requesting and obtaining LIDAR or video information from another vehicle, so it can "see" zones that are hidden from its direct vision. Potentially, the assignment of additional dedicated spectrum to ITS communications beyond the current reserved ITS band at 5.9GHz; the use of alternative spectrum (licensed or unlicensed, including mmWave frequency bands), as it is already contemplated for the NR PC5 interface; or the communications through the 5G NR Uu interface and the infrastructure, with ultra-low latency and high bandwidth, could bring many more opportunities for these use cases. 

An early work about the use of mmWave to share LIDAR data between vehicles was \cite{Perfecto2017}. This paper proposed a distributed algorithm for communication link allocation between vehicles, and concluded that it is important to consider not only link information (distance for example) but also the value of the information exchanged (receiving information from two nearby vehicles may cause resources to be expended in obtaining redundant information). However, the simulation part was focused on the scheduling algorithm and not in accurately modeling the communications part itself.

In \cite{Kutila2019}, the application of sharing LIDAR data between two vehicles was implemented in a test track with real equipment, in particular with pre-5G NR Uu interfaces, so LIDAR data was sent by one vehicle to a server in the infrastructure (traffic management center), and it was sent from there to the other vehicle. The test scenario was very demanding, since the application was the automatic driving of the second vehicle. The test results were not satisfactory, packet loss and computations can lead to large delays that make cooperative sensing only safe at very low speeds. Antenna alignments and the presence of obstacles were found to have a strong impact on performance, so the paper concluded that further optimizations are needed. 

In \cite{Convex2020}, the experiments with Rel.~14 LTE PC5 mode 4 real equipment included a see-through application to stream video from one vehicle to another. The performance was enough to send low-quality video (although not competing with other traffic). 

\textcolor{black}{See-through applications have significant bandwidth requirements. One interesting approach to meet them is to use Rel.~14 LTE PC5 mode 3 or Rel.~16 PC5 mode 1, so that direct communications between pairs of vehicles can share spectrum with communications between cellular users (vehicles or not) and the network. The network allocates resources (spectrum and power) to the different communication links, allowing spectrum re-use and optimization of different parameters (e.g., sum capacity or link delay). Although we have not found papers analyzing this configuration from the point of view of applications, the works in \cite{Aslani2020} and \cite{Guo2019b}, for example, proposed algorithms for resource allocation in Rel.~14 LTE PC5 mode 3.}

\textcolor{black}{We end the section with two tables summarizing some key points. Table~\ref{table:researchtopics_V2Vbi-directional} classifies the main contributions of the surveyed articles that study the use of V2V bi-directional communications, and Table~\ref{table:applications_V2Vbi-directional} lists specific example applications analyzed in those articles.}

\begin{table}[tb!] \footnotesize
\begin{center}
\begin{adjustbox}{max width=1.1\textwidth}
\begin{tabular}{|l|p{2.3cm}|}
\hline
\centering \textbf{Contributions} & \textbf{Related papers} \\ \hline
Distributed direct communication link allocation & \centering \arraybackslash \cite{Perfecto2017} \\ \hline
Network-based resource allocation in PC5 interface & \centering \arraybackslash \cite{Aslani2020} \cite{Guo2019b} \\ \hline 
V2V communications through the Uu interface & \centering \arraybackslash \cite{Kutila2019} \\ \hline
Demonstration/analysis of specific applications & \centering \arraybackslash \cite{Perfecto2017} \cite{Kutila2019} \cite{Convex2020} \\ \hline
Performance evaluation tested with real-life equipment & \centering \arraybackslash  \cite{Kutila2019} \cite{Convex2020} \\ \hline
\end{tabular} 
\end{adjustbox}
\caption{\textcolor{black}{Classification of contributions of surveyed papers, V2V bi-directional communications}}
\label{table:researchtopics_V2Vbi-directional}
\end{center}
\end{table}

\begin{table}[tb!] \footnotesize
\begin{center}
\begin{adjustbox}{max width=1.1\textwidth}
\begin{tabular}{|l|p{2.3cm}|}
\hline
\centering \textbf{Specific applications} & \textbf{Related papers} \\ \hline
See-through (LIDAR data) & \centering \arraybackslash \cite{Perfecto2017} \cite{Kutila2019} \\ \hline
See-through (video) & \centering \arraybackslash \cite{Convex2020} \\ \hline
\end{tabular} 
\end{adjustbox}
\caption{\textcolor{black}{Examples of specific applications analyzed on surveyed papers, V2V bi-directional communications}}
\label{table:applications_V2Vbi-directional}
\end{center}
\end{table}

\subsection{V2V multi-hop messages}

Multi-hop communications through the vehicular network using specific \mbox{C-V2X} technologies has received little attention until now, which can be due to a focus on the main case: one-hop communications. Works in \cite{Khan2018} \cite{Wang2018b} \cite{Kang2019} \cite{Alghamdi2020} proposed mechanisms to extend the coverage of safety messages, and in particular included solutions for forwarding safety messages through other vehicles, in combination with other techniques such as clustering, and relaying from the infrastructure. While the communication with the infrastructure was done with cellular networks, none of these works used specific V2V C-V2X communication technologies for the multi-hop communication part (instead, IEEE 802.11p \cite{Alghamdi2020} and generic D2D communications technologies \cite{Khan2018} \cite{Wang2018b} were used, and in \cite{Kang2019} perfect communications were assumed and the focus was on evaluating an algorithm for an optimal lane selection application).  

The work in \cite{Gupta2019b} is based on having the vehicles divided in clusters, in which vehicles can communicate directly at one hop, and two mechanisms were proposed to send event-driven safety messages to an area of interest: using vehicles at the border of two clusters to forward messages between clusters, and using relaying through the infrastructure. A key contribution, compared with the works above, is that the V2V communication technology used was Rel.~14 LTE PC5 mode 3, and a strategy for resource allocation was also proposed. 

\textcolor{black}{We end the section with two tables summarizing some key points. Table~\ref{table:researchtopics_V2Vmultihop} classifies the main contributions of the surveyed articles that study the use of V2V multi-hop messages, and Table~\ref{table:applications_V2Vmultihop} lists a specific example application analyzed in those articles.}

\begin{table}[tb!] \footnotesize
\begin{center}
\begin{adjustbox}{max width=1.1\textwidth}
\begin{tabular}{|l|p{2.3cm}|}
\hline
\centering \textbf{Contributions} & \textbf{Related papers} \\ \hline
\multirow{2}{*}{Clustering techniques} & \centering \arraybackslash \cite{Wang2018b}  \cite{Khan2018} \cite{Kang2019} \cite{Alghamdi2020} \cite{Gupta2019b} \\ \hline
Combining Uu and D2D for providing services & \centering \arraybackslash \cite{Wang2018b} \cite{Khan2018}  \cite{Alghamdi2020} \\ \hline
Network-based resource allocation in PC5 interface & \centering \arraybackslash \cite{Gupta2019b} \\ \hline
Demonstration/analysis of specific applications & \centering \arraybackslash \cite{Kang2019} \\ \hline
\end{tabular} 
\end{adjustbox}
\caption{\textcolor{black}{Classification of contributions of surveyed papers, V2V multi-hop messages}}
\label{table:researchtopics_V2Vmultihop}
\end{center}
\end{table}

\begin{table}[tb!] \footnotesize
\begin{center}
\begin{adjustbox}{max width=1.1\textwidth}
\begin{tabular}{|l|p{2.3cm}|}
\hline
\centering \textbf{Specific applications} & \textbf{Related papers} \\ \hline
Optimal lane selection & \centering \arraybackslash \cite{Kang2019} \\ \hline
\end{tabular} 
\end{adjustbox}
\caption{\textcolor{black}{Example of a specific application analyzed on surveyed papers, V2V multi-hop messages}}
\label{table:applications_V2Vmultihop}
\end{center}
\end{table}

\subsection{V2P periodic messages}

The protection of VRUs, and pedestrians in particular, is one of the most relevant use cases for safety applications for the connected vehicle. Many of the results of the works about V2V periodic messages are also relevant for V2P periodic messages, as they test the ability of the technology to enable the reception of periodic messages under timing constraints. Specific works considering V2P periodic messages with C-V2X technologies are \cite{Emara2018} \cite{Nguyen2020}, which explored the use of the LTE Uu interface to implement applications to protect pedestrians, in combination with MEC technology to reduce latency. In \cite{Emara2018}, periodic messages are sent from pedestrians' devices to vehicles through the infrastructure, so vehicles can warn drivers of risks with pedestrians when appropriate. In \cite{Nguyen2020}, periodic messages are sent from vehicles and from pedestrians' devices, and each side runs collision detection algorithms although, due to the limitations on pedestrians' devices, the paper explored the option of off-loading some of the processing to the MEC infrastructure to improve energy consumption and timing. None of the papers discuss the scalability problems posed by the fact that many pedestrians and vehicles could send periodic messages to the infrastructure that have to be broadcast back to them all. \textcolor{black}{Table~\ref{table:researchtopics_V2P} classifies the main contributions of the surveyed articles that study the use of V2P periodic messages for pedestrian protection applications.}

\begin{table}[tb!] \footnotesize
\begin{center}
\begin{adjustbox}{max width=1.1\textwidth}
\begin{tabular}{|l|p{2.3cm}|}
\hline
\centering \textbf{Contributions} & \textbf{Related papers} \\ \hline
V2P communications through the Uu interface & \centering \arraybackslash \cite{Emara2018} \cite{Nguyen2020} \\ \hline
Edge processing strategies & \centering \arraybackslash \cite{Emara2018} \cite{Nguyen2020} \\ \hline
\end{tabular} 
\end{adjustbox}
\caption{\textcolor{black}{Classification of contributions of surveyed papers, V2P periodic messages}}
\label{table:researchtopics_V2P}
\end{center}
\end{table}

\subsection{V2I periodic messages}

The straightforward implementation of this type of communication is with an RSU supporting a PC5 interface. However, some works have tried alternative implementation using the Uu interface and a V2N communication model in which the communication is IP-based from a server in the network to particular vehicles (and not to all vehicles in the coverage area of a base station/RSU). One of the challenges with the V2N implementation of these scenarios is the addressing of destination vehicles: a server can only reply after messages sent by vehicles allow to obtain their addressing information.  An example of this approach is the work in \cite{Xu2017}, where tests were done with real equipment. In the tests, messages, which could include variable traffic sign information for example, were sent periodically from a base station to vehicles through a LTE Uu interface. The latency was measured and resulted too large for safety applications. 

Experiments, in test tracks or on open roads, with real RSUs with Rel.~14 LTE PC5 mode 4 interfaces were reported in \cite{CAMP2019}, \cite{Convex2020} and \cite{Miao2021}. In \cite{CAMP2019}, tests involving up to eight cars and two RSUs were carried out, exploring different configuration options. With an appropriate configuration (HARQ on, 2RX antennas, 20 dBm transmission power, 20 MHz bandwidth), an effective coverage radio of 1400m in line-of-sight conditions was achieved; and at least 400m in non-line-of-sight conditions. The work in \cite{Convex2020} deployed an RSU in a bridge in a real road and studied the achieved radio coverage, which depended on physical conditions (obstructions) but was at least 550m with packet reception ratios over 90\% (in line with the results in \cite{CAMP2019}). Also, as previously mentioned, the work in \cite{Convex2020} reported the implementation and demonstration of several applications using real Rel.~14 LTE PC5 mode 4 equipment. One example application based on V2I periodic messages was in-vehicle information (traffic sign information sent to vehicles to be presented to drivers), which was tested in different conditions: urban and motorway scenarios, and changing locations of vehicles and RSUs. No issues were found in the behavior of the application. In \cite{Miao2021}, the focus was on applications based on V2I periodic messages for supporting automated driving, with implementations of several PC5-based C-V2X applications. The applications \textcolor{black}{were}: traffic light control system integration (an RSU broadcasts messages with traffic light information); and automated driving system integration (an RSU sends information from traffic light controllers and this information is fed to the autonomous driving module of the vehicle). The measured latency of the communication between the RSU and a vehicle was 30ms (in tests with just two vehicles and one RSU present). The importance of V2I communications to support autonomous driving vehicles was also analyzed in \cite{Guo2019}.  

\textcolor{black}{We end the section with two tables summarizing some key points. Table~\ref{table:researchtopics_V2Iperiodic} classifies the main contributions of the surveyed articles that study the use of V2I periodic messages, and Table~\ref{table:applications_V2Iperiodic} lists specific example applications analyzed in those articles.}

\begin{table}[tb!] \footnotesize
\begin{center}
\begin{adjustbox}{max width=1.1\textwidth}
\begin{tabular}{|l|p{2.3cm}|}
\hline
\centering \textbf{Contributions} & \textbf{Related papers} \\ \hline
\centering Parameter configuration of PC5 interface & \centering \arraybackslash \cite{CAMP2019} \\ \hline
V2I communications through the Uu interface & \centering \arraybackslash \cite{Xu2017} \\ \hline
Demonstration/analysis of specific applications & \centering \arraybackslash \cite{Xu2017} \cite{Convex2020} \cite{Miao2021} \\ \hline
\multirow{2}{*}{Performance evaluation tested with real-life equipment} & \centering \arraybackslash \cite{Xu2017} \cite{CAMP2019} \cite{Convex2020}  \cite{Miao2021} \\ \hline
Communications to support automatic driving & \centering \arraybackslash \cite{Guo2019} \cite{Miao2021} \\ \hline
\end{tabular} 
\end{adjustbox}
\caption{\textcolor{black}{Classification of contributions of surveyed papers, V2I periodic messages}}
\label{table:researchtopics_V2Iperiodic}
\end{center}
\end{table}

\begin{table}[tb!] \footnotesize
\begin{center}
\begin{adjustbox}{max width=1.1\textwidth}
\begin{tabular}{|l|p{2.3cm}|}
\hline
\centering \textbf{Specific applications} & \textbf{Related papers} \\ \hline
In-vehicle signage & \centering \arraybackslash \cite{Xu2017} \cite{Convex2020} \\ \hline
Green Light Optimized Speed Advisory & \centering \arraybackslash \cite{Miao2021} \\ \hline
\end{tabular} 
\end{adjustbox}
\caption{\textcolor{black}{Examples of specific applications analyzed on surveyed papers, V2I periodic messages}}
\label{table:applications_V2Iperiodic}
\end{center}
\end{table}
   
\subsection{V2I event-driven messages}

Similar to the previous case, the straightforward implementation of applications based on V2I event-driven messages is using an RSU supporting a D2D interface to broadcast information to vehicles (even to particular ones if needed). Even so, many works used the Uu interface in combination with MEC infrastructure to process and send information from a base station to vehicles with low latency. This is actually a V2N implementation of the V2I communication, but since the focus of the information sent to vehicles is local to the infrastructure station sending it, we analyze those applications here. As mentioned before, one of the challenges with the V2N implementation of these scenarios is the addressing of the destination vehicles: a server can only address vehicles after messages sent by them allow the server to obtain the destinations' addressing information and, in fact, some proposals assume that the vehicles must send periodic messages to the server.      

A key application for V2I event-driven messages is Vulnerable Road Users (VRUs) warning systems. One approach, explored in \cite{Napolitano2019}, \cite{Barmpounakis2020} and \cite{Malinverno2020}, is based on vehicles and VRUs' devices sending CAMs or BSMs to the infrastructure. The infrastructure processes the information in the CAMs/BSMs and detects risks of collisions. When a risk of collision is detected, the infrastructure sends warning messages to vehicles and VRUs. In these works the focus was on latency, and the results highlight the importance of processing the CAM information to predict collisions close to the edge of the network: perhaps using MEC in the RSU, if D2D technologies are used \cite{Barmpounakis2020}; or using MEC in base stations if Uu interfaces are used \cite{Napolitano2019} \cite{Malinverno2020}. The evaluation in \cite{Napolitano2019} and \cite{Barmpounakis2020} was done with real equipment in a lab environment, and system level simulations were carried out in \cite{Malinverno2020}.  

A similar application for V2I event-driven messages is infrastructure-based collision warning, which is intended to prevent collisions between vehicles (for example, in intersections). The work in \cite{Avino2019} designed an application with a similar approach to \cite{Napolitano2019} above: CAMs sent by vehicles are processed by the infrastructure to detect risks of collisions, and warnings are sent back to vehicles through the LTE Uu interface when needed. The results highlight again the importance of processing the CAM information to predict collisions close to the edge of the network using MEC in base stations. The work in \cite{Giannone2020} followed a similar approach: an LTE Uu interface was used to send collision warning messages from the infrastructure, but considering other network traffic as well, and ensuring low latency by using MEC and reserving some capacity in the access link for the warning messages. The evaluation in \cite{Avino2019} and \cite{Giannone2020} was done with real equipment in a lab environment. With the same communication scheme, the work in \cite{5GAA2017} investigated the importance of MEC to achieve low latency and allow heavy processing to support applications such as infrastructure-based collision warning, or infrastructure-based see-through (which can benefit from processing the video with other aggregated information in the MEC infrastructure). In a pilot test in a German motorway, a latency lower than 20 ms was achieved thanks to MEC technology \cite{5GAA2017}. 

In \cite{Sequeira2019}, vehicles send their location and movement information to a server through a 5G NR Uu interface, and the server, using machine learning, recommends merging maneuvers. 

In \cite{Tseng2020} the detection of risks of collisions was done using cameras and a computer vision algorithm. The algorithm processing was done close to the RSU that sends the warning messages, in unicast or geocast, to vehicles through Rel.~14 LTE PC5 mode 4. The focus was on exploring the parameters (transmission power, and number of repetitions of the warning messages) to achieve a given reliability requirement, but not in studying the latency of the complete system.  A vulnerable road user collision warning application based on Rel.~14 PC5 mode 4 technology was reported in \cite{Miao2021}. The application uses cameras to detect pedestrians, and an RSU broadcasts information to surrounding vehicles that determine locally if there is a risk of collision. The application in \cite{Miao2021} was implemented with real equipment with Rel.~14 LTE PC5 mode 4 interfaces (in vehicles and RSUs) and no issues were found.   

In \cite{Poli2021}, Rel.~14 LTE PC5 mode 4 technology was used to implement a cooperative lane merging application based on sending V2I event-driven messages (merging requests) combined with periodic awareness messages. An RSU sends these messages to a MEC server for processing, and the result are messages accepting or denying the merging request that are sent by the RSU to vehicles. The evaluation was done with system-level simulations. The success ratio achieved by the application was larger than 90\% if the delay in the operation of the application is very low (less than 5ms), but degrades with larger delays (delays that can be originated in the communications part or in the application information processing). The number of deployed RSUs improves reliability, but it has less impact on delay provided that there is at least one RSU per 2 km. Note that this approach was compared to using only V2V communications (see section~\ref{sec:v2vevent}) and the V2V approach was found to be better assuming enough processing capacity in vehicles.     
A simulation platform was proposed in \cite{Fu2021}, which combined vehicular simulators and a custom-made simulator of LTE PC5 technology (although, apparently, the resource reservation part was not modeled). The authors used the platform to experiment with two V2I applications for automatically driven vehicles: driving through intersections and remote driving. To support these applications, vehicles send messages to RSUs (trajectories, local sensor information) and RSUs send back driving decisions to vehicles. The RSUs communicate among themselves to be able to provide a global service to vehicles.

A different type of application was analyzed in \cite{Nkenyereye2019} and \cite{Saxena2019}, in which vehicles send messages to the infrastructure about vehicles that can be a risk to other vehicles (over-speeding vehicles, for example). The messages can be originated in an automatic system in the own vehicle that is causing the risk (for example, an over-speeding detection system), or in a nearby vehicle that detects a rogue behavior of another vehicle. After receiving the messages, the infrastructure can implement different measures, such as modifying the traffic lights in an intersection, or sending warning messages to alert other vehicles. Both D2D \cite{Nkenyereye2019} \cite{Saxena2019} and cellular (Uu based) technologies \cite{Saxena2019} were used for exchanging messages between vehicles and the infrastructure. 

As mentioned before, the work in \cite{Convex2020} implemented and demonstrated several applications in open roads and test tracks, and \textcolor{black}{used} real Rel.~14 LTE PC5 mode 4 equipment (in RSUs and vehicles). One of the applications implemented for demonstration was based on V2I event-driven messages: a roadworks warning application. No issues were found with the application in the demonstration. In \cite{Kutila2019}, a road condition warning application was deployed on a test track, but using a pre-5G NR Uu interface to send the warning messages through the infrastructure. 

\textcolor{black}{We end the section with two tables summarizing some key points. Table~\ref{table:researchtopics_V2Ieventdriven} classifies the main contributions of the surveyed articles that study the use of V2I event-driven messages, and Table~\ref{table:applications_V2Ieventdriven} lists specific example applications analyzed in those articles.}

\begin{table}[tb!] \footnotesize
\begin{center}
\begin{adjustbox}{max width=1.1\textwidth}
\begin{tabular}{|l|p{2.3cm}|}
\hline
\centering \textbf{Contributions} & \textbf{Related papers} \\ \hline
Parameter configuration of PC5 interface & \centering \arraybackslash \cite{Tseng2020} \cite{Miao2021} \\ \hline
\multirow{3}{*}{V2I communications through the Uu interface} & \centering \arraybackslash \cite{Napolitano2019} \cite{Malinverno2020} \cite{Avino2019}  \cite{Giannone2020} \cite{5GAA2017} \cite{Sequeira2019} \cite{Saxena2019} \cite{Kutila2019} \\ \hline
Combining Uu and D2D for providing V2I services &  \centering \arraybackslash \cite{Saxena2019} \\ \hline
\multirow{2}{*}{Edge processing strategies} & \centering \arraybackslash \cite{Barmpounakis2020} \cite{Napolitano2019} \cite{Malinverno2020} \cite{Avino2019} \cite{5GAA2017} \cite{Poli2021} \\ \hline
\multirow{5}{*}{Demonstration/analysis of specific applications} & \centering \arraybackslash \cite{Napolitano2019}, \cite{Barmpounakis2020} \cite{Malinverno2020} \cite{Avino2019} \cite{Giannone2020} \cite{5GAA2017}  \cite{Sequeira2019} \cite{Tseng2020} \cite{Miao2021} \cite{Poli2021} \cite{Fu2021} \cite{Nkenyereye2019} \cite{Saxena2019}  \cite{Convex2020} \cite{Kutila2019} \\ \hline
\multirow{3}{*}{Performance evaluation tested with real-life equipment} & \centering \arraybackslash \cite{Napolitano2019} \cite{Barmpounakis2020}  \cite{Avino2019} \cite{Giannone2020} \cite{5GAA2017} \cite{Miao2021}  \cite{Convex2020} \cite{Kutila2019} \\ \hline
Communications to support automatic driving & \centering \arraybackslash \cite{Fu2021} \\ \hline
\end{tabular} 
\end{adjustbox}
\caption{\textcolor{black}{Classification of contributions of surveyed papers, V2I event-driven messages}}
\label{table:researchtopics_V2Ieventdriven}
\end{center}
\end{table}

\begin{table}[tb!] \footnotesize
\begin{center}
\begin{adjustbox}{max width=1.1\textwidth}
\begin{tabular}{|l|p{3cm}|}
\hline
\centering \textbf{Specific applications} & \textbf{Related papers} \\ \hline
Vulnerable road user collision warning & \centering \arraybackslash \cite{Napolitano2019}, \cite{Barmpounakis2020} \cite{Malinverno2020} \cite{Miao2021} \\ \hline
Infrastructure-based collision warning & \centering \arraybackslash \cite{5GAA2017} \cite{Avino2019} \cite{Giannone2020} \cite{Tseng2020} \\ \hline
Infrastructure-based see-through & \centering \arraybackslash \cite{5GAA2017} \\ \hline
Cooperative lane merging & \centering \arraybackslash \cite{Sequeira2019} \cite{Poli2021} \\ \hline
Driving through intersections support & \centering \arraybackslash \cite{Fu2021} \\ \hline
Warning of rogue vehicles & \centering \arraybackslash \cite{Nkenyereye2019} \cite{Saxena2019} \\ \hline
Roadworks warning & \centering \arraybackslash \cite{Convex2020} \\ \hline
Road condition warning & \centering \arraybackslash \cite{Kutila2019} \\ \hline
\end{tabular} 
\end{adjustbox}
\caption{\textcolor{black}{Examples of specific applications analyzed on surveyed papers, V2I event-driven messages}}
\label{table:applications_V2Ieventdriven}
\end{center}
\end{table}
 
\subsection{RSU-relayed messages}

\textcolor{black}{The solution explored} in \cite{Gupta2018} \cite{Gupta2019} \cite{Gupta2019b} \textcolor{black}{organized} vehicles in clusters, with periodic and event-driven V2V messages sent to the infrastructure in order to be relayed to other clusters through their respective cluster heads. The Uu interface \textcolor{black}{was} used for these communications, which means that this implementation really used V2N communications. These works are a proof-of-concept of techniques to increase the coverage of ITS messages, but the approach is less efficient than using an RSU that supports a D2D interface and can directly receive V2V messages exchanged by vehicles. For example, in \cite{Alghamdi2020} \cite{Guo2019} \cite{Barmpounakis2020}, the information obtained from V2V periodic messages received by RSUs \textcolor{black}{was} relayed to servers for processing (for gathering information \cite{Alghamdi2020} \cite{Guo2019} or detecting risks of collisions \cite{Barmpounakis2020}). In \cite{Nkenyereye2019}, event-driven messages with traffic violation reports (e.g., over-speeding notifications) transmitted by vehicles were sent to a traffic control center by RSUs, which also broadcast them to increase coverage, perhaps after some processing such as aggregating the information of several messages. In \cite{Chang2020}, shockwave detection algorithms were studied, and the solution was based on having a network that allows the infrastructure to receive messages sent by vehicles with their location and movement information, which are then relayed to servers for running the algorithms with the information of the messages as input. In principle, implementations in which the information is directly sent by the vehicles to the servers and implementations in which RSUs relay the messages to the servers are both possible. When processing of the received information is required, the solutions use MEC to keep the processing close to the edge of the network and achieve shorter latency \cite{Barmpounakis2020} \cite{Nkenyereye2019} \cite{Chang2020}, while relying on computations in the cloud for operations that require a global view of the network \cite{Barmpounakis2020} \cite{Chang2020}. 

\textcolor{black}{We end the section with two tables summarizing some key points. Table~\ref{table:researchtopics_RSUrelayed} classifies the main contributions of the surveyed articles that study the use of RSU-relayed messages, and Table~\ref{table:applications_RSUrelayed} lists specific example applications analyzed in those articles.}

\begin{table}[tb!] \footnotesize
\begin{center}
\begin{adjustbox}{max width=1.1\textwidth}
\begin{tabular}{|l|p{2.3cm}|}
\hline
\centering \textbf{Contributions} & \textbf{Related papers} \\ \hline
Clustering techniques & \centering \arraybackslash \cite{Gupta2018} \cite{Gupta2019} \cite{Gupta2019b} \\ \hline
Relaying through the Uu interface & \centering \arraybackslash  \cite{Gupta2018} \cite{Gupta2019} \cite{Gupta2019b} \\ \hline
Edge processing strategies & \centering \arraybackslash \cite{Barmpounakis2020} \cite{Nkenyereye2019} \cite{Chang2020} \\ \hline
\multirow{2}{*}{Demonstration/analysis of specific applications} & \centering \arraybackslash \cite{Alghamdi2020} \cite{Guo2019} \cite{Barmpounakis2020} \cite{Nkenyereye2019} \cite{Chang2020} \\ \hline
\end{tabular} 
\end{adjustbox}
\caption{\textcolor{black}{Classification of contributions of surveyed papers, RSU-relayed messages}}
\label{table:researchtopics_RSUrelayed}
\end{center}
\end{table}

\begin{table}[tb!] \footnotesize
\begin{center}
\begin{adjustbox}{max width=1.1\textwidth}
\begin{tabular}{|l|p{2.3cm}|}
\hline
\centering \textbf{Specific applications} & \textbf{Related papers} \\ \hline
Relaying information to Traffic Control Center & \centering \arraybackslash \cite{Alghamdi2020} \cite{Guo2019} \cite{Nkenyereye2019} \\ \hline
Collision avoidance & \centering \arraybackslash \cite{Barmpounakis2020} \cite{Nkenyereye2019} \\ \hline
Shockwave detection & \centering \arraybackslash \cite{Chang2020} \\ \hline
\end{tabular} 
\end{adjustbox}
\caption{\textcolor{black}{Examples of specific applications analyzed on surveyed papers, RSU-relayed messages}}
\label{table:applications_RSUrelayed}
\end{center}
\end{table}

\subsection{V2N \textcolor{black}{bi-directional} communications}

In \cite{Xu2017}, using real equipment on a test track, media downloading (maps, for instance) through a LTE Uu interface was compared with using IEEE 802.11p, and the achieved throughput was clearly better in the former case. It is interesting that, in \cite{Xu2017}, the performance of the LTE Uu interface was tested with different vehicles' speeds and a significant degradation was observed for speeds over 90 km/h. 

The works in \cite{Khan2018} (simulations) and \cite{Convex2020} (field tests) explored the ability of network slicing techniques to offer different service quality to different applications. In particular, in \cite{Convex2020}, LTE Uu based V2N communications from devices located in a vehicle moving through a motorway, with coverage provided by a test cellular network, were evaluated. The results of using network slicing techniques showed that the network is able to offer different services (bandwidth and latency) to different users. Additionally, the results showed that when one slice does not use its assigned bandwidth, it can be used by other slices, proving that network slicing is able to guarantee resources to one application, without hindering the efficient use of resources. 

Several works \cite{Wang2018b} \cite{Ding2019} proposed a model in which cellular technology is used to send information to a traffic server in the network (e.g., a traffic control centre) and retrieve back traffic incident notifications, or path planning data. In \cite{Ding2019}, vehicles, through 5G NR Uu interfaces, send cooperative perception messages and download local dynamics maps to support automatic driving, with processing done in a MEC infrastructure to achieve the required latency. The evaluation was based on simulations. In \cite{Li2019}, the target application was dynamic map downloading, but the radio technology was mmWave, and SDN was used to configure the mmWave links and the network. The evaluation was done in a testbed with one vehicle and two RSUs.           

\textcolor{black}{We end the section with two tables summarizing some key points. Table~\ref{table:researchtopics_V2N} classifies the main contributions of the surveyed articles that study the use of V2N bi-directional communications, and Table~\ref{table:applications_V2N} lists a specific example application analyzed in those articles.}

\begin{table}[tb!] \footnotesize
\begin{center}
\begin{adjustbox}{max width=1.1\textwidth}
\begin{tabular}{|l|p{2.3cm}|}
\hline
\centering \textbf{Contributions} & \textbf{Related papers} \\ \hline
\multirow{2}{*}{V2N communications through the Uu interface} & \centering \arraybackslash \cite{Xu2017} \cite{Khan2018} \cite{Convex2020} \cite{Wang2018b} \cite{Ding2019} \\ \hline
Edge processing strategies & \centering \arraybackslash \cite{Ding2019} \\ \hline
Comparison of IEEE 802.11p with C-V2X & \centering \arraybackslash \cite{Xu2017} \\ \hline
Differentiated quality of service & \centering \arraybackslash \cite{Khan2018} \cite{Convex2020} \\ \hline
Demonstration/analysis of specific applications & \centering \arraybackslash \cite{Wang2018b} \cite{Ding2019} \cite{Li2019} \\ \hline
Performance evaluation tested with real-life equipment & \centering \arraybackslash \cite{Xu2017} \cite{Convex2020} \cite{Li2019} \\ \hline
Communications to support automatic driving & \centering \arraybackslash \cite{Ding2019} \\ \hline
\end{tabular} 
\end{adjustbox}
\caption{\textcolor{black}{Classification of contributions of surveyed papers, V2N bi-directional communications}}
\label{table:researchtopics_V2N}
\end{center}
\end{table}

\begin{table}[tb!] \footnotesize
\begin{center}
\begin{adjustbox}{max width=1.1\textwidth}
\begin{tabular}{|l|p{2.3cm}|}
\hline
\centering \textbf{Specific applications} & \textbf{Related papers} \\ \hline
Dynamic map download and update & \centering \arraybackslash \cite{Wang2018b} \cite{Ding2019} \cite{Li2019} \\ \hline
\end{tabular} 
\end{adjustbox}
\caption{\textcolor{black}{Example of a specific application analyzed on surveyed papers, V2N bi-directional communications}}
\label{table:applications_V2N}
\end{center}
\end{table}

\section{Challenges for C-V2X applications research and deployment}
\label{sec:challenges}
\textcolor{black}{
This section summarizes the lessons learnt from our survey and, based on them, identifies research challenges in the use of C-V2X technologies to develop applications for road safety and traffic efficiency. We can outline the following key findings from our survey:}
\textcolor{black}{
\begin{enumerate}
\item Many works confirm the feasibility of using C-V2X technologies to implement road safety and traffic efficiency applications. Different communication models and C-V2X technologies have been explored to implement specific applications (in simulations or with real equipment), showing that they can perform properly.  
\item The type of communication that has received the most attention is V2V using periodic messages. On the other hand, V2P periodic messages and V2V multi-hop stand out for the small number of works covering these types of communications applied to road safety and traffic efficiency applications based on C-V2X technologies. 
\item In applications that communicate through the C-V2X PC5 interface, a degradation in performance when network load increases has been identified \cite{Molina-Masegosa2018} \cite{Sabeeh2019} \cite{Hirai2020}, but only a few works study congestion control mechanisms to deal with it. 
\item A related topic is resource allocation for the PC5 interface. The distributed resource allocation algorithm (S-SPS) specified by 3GPP has been thoroughly studied in the context of supporting applications, and its weaknesses (half-duplex problem, bursts of collisions, and dealing with non-periodic or variable size messages) have been identified, with some proposals for improvement. The modifications to the algorithm in Rel.~16 still need detailed evaluation. On the other hand, in the case of resource allocation from the network, 3GPP does not define a particular allocation algorithm, and different works have proposed optimization techniques to allocate resources to different links while maximizing some properties (such as sum capacity).  
\item Few works have evaluated NR technologies (NR PC5 mode 1, NR PC5 mode 2, or NR Uu). These technologies are the most appropriate to implement advanced road safety applications with high bandwidth demands. Additionally, we are still in the early stages of discovering the possibilities of applying new network paradigms such as SDN, network virtualization and network slicing in the context of road safety and traffic efficiency applications. On the other hand, the ability of MEC technology to decrease the latency in V2I and V2N based services has been confirmed in different works.   
\item Different evaluation methods have been used in the surveyed papers: analytical evaluation, and experiments using simulations and tests with real equipment in test tracks and open roads. On the one hand, this shows the progress of the research on \mbox{C-V2X} technologies. On the other hand, there is still relatively few works reporting pilot deployments, specially compared with IEEE 802.11p-based solutions and, for the most part, simulations in the surveyed literature are based on proprietary software or custom-made modules, even for implementing common standardized functionalities.  
\end{enumerate} 
}
\textcolor{black}{
Considering these key points and some missing ones in the literature, we identify the technological and practical challenges for using C-V2X technologies to deploy road safety and traffic efficiency applications for the connected vehicle in the real world. We highlight the following (summarized in Table~\ref{table:challenges}):}

\begin{table}[tb!] \footnotesize
\begin{center}
\begin{adjustbox}{max width=0.85\textwidth}
\begin{tabular}{|p{4.5cm}|p{6cm}|}
\hline
\centering \textbf{Topic} & \centering \arraybackslash \textbf{Challenge} \\ \hline
\vspace{-0.1cm} Congestion control and distributed resource allocation &  Algorithms that integrate congestion control and distributed resource allocation in an efficient manner \\ \hline
Network-based resource allocation and advanced safety applications & Developing new algorithms for bandwidth-hungry applications \\ \hline
\multirow{2}{*}{Geobroadcast} & Evaluation of geobradcast routing algorithms with C-V2X PC5 technologies \\ \hline
\multirow{3}{*}{Advanced applications} & Developing applications over NR PC5 and NR Uu interfaces. Explore the role of SDN, network virtualization and network slicing  \\ \hline 
\multirow{3}{*}{Multi-operator scenarios} & Interoperability and resource sharing when users of different operators communicate through an application\\ \hline 
\multirow{2}{*}{Experimentation - real equipment} & Increase the number of test pilots and real life demonstration of applications \\ \hline 
\multirow{4}{*}{Experimentation - simulation tools} & Adopt and support common set of simulation tools that implement the complete set of C-V2X technologies and the rest of the protocol stack up to the application layer \\ \hline
\multirow{4}{*}{Hybrid solutions} & Compatibility approaches and the study of the benefits of combining the service of different interfaces (IEEE 802.11p, LTE PC5, NR PC5 and NR Uu). \\ \hline
\multirow{2}{*}{Business models} & Design business models to drive the deployment of vehicular networks \\ \hline
\end{tabular} 
\end{adjustbox}
\caption{\textcolor{black}{Research challenges in C-V2X based road safety and traffic efficiency applications}}
\label{table:challenges}
\end{center}
\end{table}

\textcolor{black}{
\begin{enumerate}
\item \textbf{Congestion control and distributed resource allocation:} The performance of the vehicular network must be robust \textcolor{black}{regardless} of network load but, since different situations can involve very different numbers of vehicles and other actors (pedestrians, RSUs), the network load cannot be anticipated, and a congestion control mechanism to deal with situations in which the offered load degrades network performance is needed. There are several proposals, such as using the congestion control algorithms defined for IEEE 802.11p-based protocol stacks \cite{5GAA2019} \cite{Shimizu2020}, or defining new ones \cite{Choi2020}. ETSI has standardized a solution \cite{ETSI2018} for congestion control in the PC5 interface, but it is a simple one based on states, an approach that has been proved inferior to linear control systems in the extensive literature on the topic in IEEE 802.11p-based approaches \cite{Bansal2013} \cite{Amador2020}. The interaction of the congestion control mechanism with a scheduling based on semi-persistent assignment of resources, which introduces a delay to the adaptation of message rates, requires further investigation. Additionally, the relation with the HARQ mechanism used in the PC5 interface, which sends each packet twice (or even more in the case of NR PC5) to increase reliability, has to be evaluated. 
\item \textcolor{black}{\textbf{Network-based resource allocation and advanced safety applications:} Advanced  applications for road safety and traffic efficiency may require large spectrum resources to support point-to-point bi-directional links between pairs of vehicles, and between vehicles and the network. There can be scenarios with dedicated spectrum for advanced vehicular applications, or scenarios where these applications share the spectrum with cellular users. In both cases, network-based allocation of resources enables the efficient sharing of resources, but there is the need to develop resource allocation algorithms optimized for the requirements of different advanced vehicular applications.}
\item \textbf{Geobroadcast:} Different applications for the connected vehicle are based on extending the coverage of ITS messages by forwarding them in an geographical area of interest, potentially over multiple hops across vehicles. For example, this approach can be used to warn vehicles of an incident in a motorway and give them more time to react or even take a detour, or it can be used in a green light optimized speed advisory system to provide more time to adapt the speed of vehicles to avoid having to stop at the traffic light. This support of multiple-hop communications can be provided by forwarding algorithms that have already been studied in the context of IEEE 802.11p-based protocol stacks, with solutions such as the Greedy forwarding and Contention-Based Forwarding (CBF) algorithms \cite{ETSI2020}. These solutions could be applied to D2D PC5 technology, but this has to be carefully analyzed because the particularities of the S-SPS algorithm, such as the granularity of the possible times between consecutive messages or the introduced delay, which may have important implications to the performance of CBF in particular. On the other hand, there are opportunities to complement the forwarding of messages through other vehicles with relying messages through the infrastructure. Message relaying is, in fact, a topic under study in future 3GPP releases (Release 17).
\item \textbf{Advanced applications:} One of the objectives of C-V2X is to enable services beyond the basic ones considered in 802.11p-based networks. The combination of D2D communication services between vehicles with the ability of 5G networks to fulfill both ultra-low latency requirements and high bandwidth demands opens up possibilities for applications such as see-through, remote driving, massive sharing of sensor information, and many others. However, much more work is needed to explore how to use the most recent C-V2X solutions and related technologies to implement these applications (5G NR mode 1, 5G NR mode 2, 5G NR Uu, MEC, SDN, network virtualization and network slicing) since most works in the C-V2X literature are still dealing with basic applications. 
\item \textbf{Multi-operator scenarios}: Applications for the connected vehicle have to support interoperability for users from different network operators, while keeping the strict latency requirements of safety applications. This could be challenging, in particular when processing is done at the edge of an operator network to improve performance. In D2D technologies with network assistance, the assignment of resources from base stations of different operators has to be coordinated. Even in D2D communications without network assistance, some operation parameters are configured in advance when the device is connected to the network. A solution would require coordination between operators and cells. For example, spectrum management seems particularly challenging.   
\item \textbf{Experimentation:} While a wide variety of evaluation tools are available for IEEE 802.11p-based vehicular networks, this has not been the case so far for C-V2X technologies, although this situation is improving rapidly. For example, low-priced IEEE 802.11p cards are available, allowing experiments with real equipment. Recently, commercial equipment supporting C-V2X technologies, in particular LTE PC5 mode 4, is being made available \cite{5GAA2020}, although at a cost that can be an order of magnitude higher than IEEE 802.11p equipment. As for simulation tools, the use of many custom-made and proprietary software that we have observed in the evaluation part of works in the literature makes more difficult to compare the  different results and may hinder efforts to replicate results. \textcolor{black}{There are established open source simulators that implement IEEE 802.11p-based protocol stacks such as Artery~\cite{Riebl2015} and Veins~\cite{VeinsLTE2014}), which also implement scenarios with D2D C-V2X communications using OMNET++ and SimuLTE~\cite{SimuLTE2015}. A more recent work includes an extension to Artery and Veins, called Open C-V2X~\cite{mccarthy2021opencv2x}, which takes advantage of the Artery implementation of the ETSI ITS stack and adds D2D capabilities including a Rel.~14 Mode 4 scenario. \mbox{Artery-C}~\cite{ArteryCFestag} is another Artery-based toolkit that has recently come to light which the authors intend to publish as open source. Moreover, there are simulation tools based on ns-3 (e.g., ns-3\_c-v2x~\cite{Eckermann2019cv2xsim}) that model C-V2X Mode 4, however, the aforementioned solution focuses more on the lower layers rather than on the Facilities or Application layers (e.g., ns3\_c-v2x only sends dummy packets at fixed rates to simulate CAMs or BSMs). Outside of the open-source community, there are simulation tools (e.g., LTEV2Vsim~\cite{ltev2vsim}) that use Matlab, which also provides an LTE toolbox which supports V2X and sidelink direct communications.} Therefore, the tools are beginning to be available, and the research and open-source community has to support and use established simulation tools, which would make easier the exchange and replicability of results. This point is even more challenging due to the evolution of C-V2X technologies (such as 5G NR PC5), so there is a need to implement these new technologies in real equipment and simulators and make them available to the research community. 
\item \textbf{Hybrid solutions}: The combination of IEEE 802.11p for V2V communications and LTE Uu for V2I and V2N communications has received some attention in the literature \cite{Zheng2015} \cite{Abboud2016} \cite{Ucar2016} \cite{Qi2020}. However, the role of networks based on IEEE 802.11p and their potential combination with C-V2X networks through the PC5 and Uu interface is still an open issue, with some initial works exploring architectures for heterogeneous access networks \cite{Hameed2020} \cite{Jacob2020} and others looking at the co-existence of IEEE 802.11p and LTE PC5 mode 4 in the same radio channel \cite{Roux2020}. Even within \mbox{C-V2X} technologies, the role of the different interfaces --- LTE PC5, NR PC5, and NR Uu --- in the provision of services is also an open issue. More research is needed on compatibility approaches and on how specific applications or set of applications could benefit from combining the service provided by different interfaces.   
\item \textbf{Business models:} There is a great challenge in investigating business models that can drive the deployment of vehicular networks \cite{Usman2020}. While Business-to-Consumer (B2C) models are challenging from the business side (e.g., resource consumption from cloud-based V2P safety applications, multi-operator scenarios) and the consumer side (e.g., privacy concerns, data and battery consumption), Business-to-Business (B2B) models can provide a niche for operators, vehicle manufacturers, and government agencies (e.g., the X as a Service (aaS) business model). It is believed there might be players that would more clearly have business opportunities in the context of C-V2X technologies, but this requires further investigation, particularly with D2D technologies. 
\end{enumerate}
}

\section{Conclusion}
\label{sec:conclusion}
This article reviews the literature on the use of C-V2X technologies to develop road safety and traffic efficiency applications for connected vehicles. After describing the relevant C-V2X communication technologies, we have proposed a classification for the different types of communication needed to support applications for connected vehicles. Following that classification, we have described how the research community has analyzed and validated the use of C-V2X technologies specified by 3GPP to develop road safety and traffic efficiency applications, exploring its limitations, and proposing improvements that have contributed to specifications development. These works have shown the different ways in which C-V2X communications technologies can be used to implement communication-based applications in vehicles. However, this survey also identifies the need for more work in several fields of research: the C-V2X technologies are evolving and there is the need to explore the possibilities of their most recent versions (e.g., 5G NR PC5 or NR Uu in combination with technologies such as MEC); the development and validation of advanced services; the combination of different technologies to provide V2X services; the interoperability in multi-operator scenarios; and the business and deployment models that can drive the adoption of the C-V2X technology in the real world.   

\bibliography{review_vehicular_applications_5G}

\end{document}